\documentclass[aps,prb,twocolumn,groupedaddress,floatfix]{revtex4-1}
\usepackage[dvips]{graphicx}
\begin{document}

\renewcommand{\vec}[1]{\mathbf{#1}}




\title{\textit{Ab initio} lattice dynamical studies of silicon clathrate frameworks and their negative thermal expansion}


\author{Ville J. H\"{a}rk\"{o}nen}

\author{Antti J. Karttunen}
\email[]{antti.j.karttunen@iki.fi}

\affiliation{University of Jyv\"{a}skyl\"{a}, PO Box 35, FI-40014, Finland}


\date{\today}

\begin{abstract}
The thermal and lattice dynamical properties of seven silicon clathrate framework structures are investigated with \textit{ab initio} density functional methods (frameworks I, II, IV, V, VII, VIII, and H). The negative thermal expansion (NTE) phenomenon is investigated by means of quasiharmonic approximation and applying it to equal time displacement correlation functions. The thermal properties of the studied clathrate frameworks, excluding the VII framework, resemble those of the crystalline silicon diamond structure. The clathrate framework VII was found to have anomalous NTE temperature range up to 300 K and it is suitable for further studies of the mechanisms of NTE. Investigation of the displacement correlation functions revealed that in NTE, the volume derivatives of the mean square displacements and mean square relative displacements of atoms behave similarly to the vibrational entropy volume derivatives and consequently to the coefficients of thermal expansion as a function of temperature. All studied clathrate frameworks, excluding the VII framework, possess a phonon band gap or even two in the case of the framework V.
\end{abstract}

\pacs{63.20.dk, 63.70.+h, 65.40.De, 65.60.+a, 65.80.-g}
\keywords{negative thermal expansion, phonon dispersion, clathrate, \textit{ab initio} lattice dynamics, DFTP, displacement correlation function, momentum correlation function, mean square displacement, vibrational entropy, silicon.}

\maketitle

\section{Introduction}
\label{cha:Introduction}
 In addition to the crystalline silicon diamond structure (\textit{d}-Si), various other crystal structures contain tetrahedrally bonded silicon atoms. Examples of such structures are the experimentally known semiconducting clathrates,\cite{Kasper24121965.science.clath.1965} which are typically obtained as host-guest compounds such as $\textrm{Na}_{24}\textrm{Si}_{136}$, where a porous Si clathrate II framework (sometimes denoted as Si$_{34}$ or Si$_{136}$) is partially filled by Na guest atoms.\cite{Shevelkov-2011-zintl-clath} However, silicon clathrates with a (practically) guest-free Si framework structure have been prepared, as well.\cite{PhysRevB.62.R7707.empty.clathrates.Gryko,Ammar2004393-clath} The semiconducting clathrates have been investigated intensively due to their interesting properties such as anomalous electronic structure (differing from that of \textit{d}-Si),\cite{Adams.et.al.1994-PhysRevB.49.8048,Saito.et.al.electr.calc.clathI-PhysRevB.51.2628-1995} superconductivity\cite{PhysRevLett.74.1427.Si.clath.superconductivity} and relatively high thermoelectric efficiency.\cite{4899409Nolas-1998-Ge-clath-thermoelect.,Sootsman-thermoelect.-review.2009-ANIE200900598,Christensen.et.al.-2010-B916400F} In particular, the most intense research efforts on the thermoelectric properties of clathrates have so far been directed at germanium-based clathrates such as $\textrm{Ba}_{8}\textrm{Ga}_{16}\textrm{Ge}_{30}$.\cite{Kuznetsov-et.al.2000-jap/87/11/10.1063/1.373469,Toberer.et.al-2008-PhysRevB.77.075203}

Tang \textit{et al.} have investigated the thermal properties of the clathrate II framework by means of \textit{ab initio} calculations and experimental measurements and have shown that the clathrate II framework exhibits negative thermal expansion (NTE) in the temperature range of 10--140 K.\cite{PhysRevB.74.014109.Tang.et.al.thermal.prop.clathII} NTE is a phenomenon in which a material contracts instead of expanding when heated to higher temperatures. From the point of view of practical applications, NTE materials have been used to design composite materials with zero thermal expansion.\cite{Evans1997311-NTE-1998} The NTE materials have been studied actively after the discovery of large temperature range NTE materials such as $\textrm{ZrW}_{2}\textrm{O}_{8}$.\cite{chem.mat.9602959.Evans.Mary.1996.NTE,Mary05041996.science.ZrW2O8.NTE}, which shows NTE in a wide temperature range from nearly 0 K up to 1080 K. The underlying mechanisms causing the NTE behavior are not fully understood. The NTE phenomenon and the different NTE mechanisms have been reviewed for example by Barrera \textit{et al.}\cite{0953-8984-17-4-R03.jour.phys.cond.mat..Barrera.Bruno.Barron.NTE-rew.} \textit{d}-Si also shows NTE within temperature range from 25 K up to 120 K.\cite{Reeber1996259.Sialpha.thermalexp.experimental} However, to our knowledge, there are no elemental semiconductors with anomalously large NTE temperature ranges comparable to the oxides such as $\textrm{ZrW}_{2}\textrm{O}_{8}$.

The NTE behavior of materials can be experimentally studied by means of Extended X-Ray Absorption Fine Structure (EXAFS) spectroscopy, which enables the measurement of local dynamics of the crystal lattice.\cite{PhysRevLett.27.1204.Sayers.et.al.1971.EXAFS,PhysRevB.11.2795.Lee.et.al.EXAFS.1975,0022-3719-9-24-006.Hayes.et.al.EXAFS1976,PhysRevB.49.888.Troger.et.al.EXAFS.1994} Quantities which can be obtained from EXAFS spectra are, for example, mean square displacements (MSD) and mean square relative displacements (MSRD). Consequently, EXAFS has also been used to study the NTE phenomenon.\cite{PhysRevB.73.214305.Sanson.et.al.EXAFS.NTE.2006,0953-8984-24-11-115403.Abd.et.al.EXAFS.NTE.2012} However, computational studies of the NTE phenomenon from the point of view of MSRD are not common. To our knowledge, there are no previous studies where the NTE behavior of materials is investigated by means of full \textit{ab initio} calculation of MSRDs.  

In this paper we investigate the thermal properties and negative thermal expansion of seven different silicon clathrate frameworks by means of \textit{ab initio} lattice dynamics. The thermal properties of the materials are determined within quasiharmonic approximation (QHA).\cite{born-huang1954dynamical} The results are compared with available experimental data and previous computational predictions for \textit{d}-Si to establish the accuracy of the computational methods. We use equal time displacement correlation functions to investigate the mechanisms of the thermal expansion and to elucidate the negative thermal expansion behavior of the silicon clathrate frameworks.

\section{Theory, computational methods and studied structures}
\label{cha:Theory calc.meth.}
\subsection{Lattice dynamics and thermal expansion}
\label{L.dyn.therm.exp.}
To obtain the various thermal properties, the phonon eigenvalues and phonon eigenvectors have been extracted with \textit{ab initio} methods, by applying density functional perturbation theory (DFPT).\cite{RevModPhys.73.515-Baroni-2001-DFTP} The DFTP formalism enables the calculation of dynamical matrix (introduced by the relations below), after which the phonon eigenvalues and eigenvectors can be solved from an eigenvalue equation written within the harmonic approximation as\cite{maradudin1971-harm-appr}
\begin{equation}
\omega^{2}_{j}\left(\vec{q}\right)e_{\alpha}\left(\kappa|\vec{q}j\right) = \sum_{\kappa',\beta}D_{\alpha\beta}\left(\kappa\kappa'|\vec{q}\right)e_{\beta}\left(\kappa'|\vec{q}j\right),
\label{eq:eigenvalueeq}   
\end{equation}
where $\alpha$ and $\beta$ are cartesian indices, $\kappa$ and $\kappa'$ are atom indices within the primitive unit cell, $\omega_{j}\left(\vec{q}\right)$ is phonon eigenvalue (referred as phonon frequency from now on) for phonon branch $j$ and wave vector $\vec{q}$, $e_{\alpha}\left(\kappa|\vec{q}j\right)$ are the components of the phonon eigenvector and  $D_{\alpha\beta}\left(\kappa\kappa'|\vec{q}\right)$ are the components of the dynamical matrix. The dynamical matrix is the Fourier transform of the real space force constant matrix $\Phi_{\alpha\beta}\left(l\kappa;l^{,}\kappa^{,}\right)$
\begin{eqnarray}
D_{\alpha\beta}\left(\kappa\kappa'|\vec{q}\right)=\left(M_{\kappa}M_{\kappa'}\right)^{-1/2}\sum_{l'}\Phi_{\alpha\beta}\left(l\kappa;l'\kappa'\right)
\nonumber\\
\times{}e^{-i\vec{q}\cdot\left[\vec{x}\left(l\right)-\vec{x}\left(l'\right)\right]},
\end{eqnarray}
where  $M_{\kappa}$ and $M_{\kappa'}$ are the atomic masses of atoms $\kappa$ and $\kappa'$ respectively, and $\vec{x}\left(l\right)$ is the position vector of the $l_{\textrm{th}}$ unit cell. The components of the eigenvector $\vec{e}\left(\kappa|\vec{q}j\right)$ are chosen to satisfy the orthonormality and closure conditions\cite{born-huang1954dynamical}
\begin{equation}
\sum_{\kappa,\alpha}e_{\alpha}\left(\kappa|\vec{q}j'\right)e^{*}_{\alpha}\left(\kappa|\vec{q}j\right)=\delta_{jj'},
\label{eq:orthonorm}
\end{equation}
\begin{equation}
\sum_{j}e_{\alpha}\left(\kappa|\vec{q}j\right)e^{*}_{\beta}\left(\kappa'|\vec{q}j\right)=\delta_{\alpha\beta}\delta_{\kappa\kappa'},
\label{eq:closure}
\end{equation}
where $\delta_{\alpha\beta}$ is the Kronecker delta. The volumetric coefficient of thermal expansion (CTE) $\alpha_{V}$ is defined as
\begin{equation} \alpha_{V}=\frac{1}{V_{0}}\frac{\partial{V}}{\partial{T}}, \end{equation}
where $V$ is the volume, $V_{0}$ is the equilibrium volume and $T$ the temperature. The linear CTE is a measure of the change in particular dimension of the crystal with respect to temperature. In the cubic crystal system, linear CTE $\alpha_{L}$ in the direction of the principal axis can be written as
\begin{equation} 
\alpha_{L}=\frac{1}{L_{0}}\frac{\partial{L}}{\partial{T}}=\frac{1}{3V_{0}}\frac{\partial{V}}{\partial{T}}=\frac{1}{3}\alpha_{V}.
\label{eq:CTEL}
\end{equation}
In non-cubic crystals, the anisotropy of the crystal structure must be taken in account. That is, the different components of the thermal expansion along different principal axes must be calculated separately to obtain the linear CTEs. In this work only volumetric CTEs have been calculated for non-cubic structures. Using QHA, the expression for the volumetric CTE can be written in terms of isothermal bulk modulus $B$, phonon mode Gr\"{u}neisen parameters $\gamma_{j}\left(\vec{q}\right)$ and phonon mode heat capacities at constant volume $c_{v,j}\left(\vec{q}\right)$\cite{ashcroft-solid-state-phys}
 \begin{equation}
\alpha_{V}=\frac{1}{B}\sum_{\vec{q},j}c_{v,j}\left(\vec{q}\right)\gamma_{j}\left(\vec{q}\right),
\label{eq:CTEV2}
\end{equation}
The heat capacity can be expressed in terms of Bose-Einstein distribution function $\overline{n}_{\vec{q},j}$ and phonon frequencies $\omega_{j}\left(\vec{q}\right)$
 \begin{equation}
c_{v,j}\left(\vec{q}\right)=\frac{\omega_{j}\left(\vec{q}\right)}{V_{0}}\frac{\partial}{\partial{T}}\overline{n}_{\vec{q},j},
\label{eq:heatcapacity}
\end{equation}
and the Gr\"{u}neisen parameters can be expressed as
\begin{equation}
\gamma_{j}\left(\vec{q}\right)=-\frac{V_{0}}{\omega_{0,j}\left(\vec{q}\right)}\frac{\partial{\omega_{j}}\left(\vec{q}\right)}{\partial{V}},
\label{eq:gruneisenparameter}
\end{equation}
where $\omega_{0,j}\left(\vec{q}\right)$ is the phonon frequency for the phonon mode $j$ and wave vector $\vec{q}$ at the equilibrium volume $V_{0}$. The Gr\"{u}neisen parameters can also be expressed as\cite{taylor1998thermal}
\begin{eqnarray}
\gamma_{j}\left(\vec{q}\right) = -\frac{1}{2}\frac{V_{0}}{\omega^{2}_{0,j}\left(\vec{q}\right)}\sum_{\alpha,\beta}\sum_{\kappa,\kappa'}e^{*}_{\alpha}\left(\kappa|\vec{q}j\right)
 \nonumber\\
\times{}\frac{\partial D_{\alpha\beta}\left(\kappa\kappa'|\vec{q}\right)}{\partial{V}}e_{\beta}\left(\kappa'|\vec{q}j\right).
\label{eq:gruneisenparameter2}
\end{eqnarray}
Eq. \ref{eq:CTEV2} shows that when NTE occurs, the Gr\"{u}neisen parameters must have negative values since the bulk modulus and the heat capacities have positive values. Furthermore, from Eq. \ref{eq:gruneisenparameter} it can be seen that to have negative Gr\"{u}neisen parameter values, the phonon frequencies must decrease with decreasing volume. Analogously, Eq. \ref{eq:gruneisenparameter2} implies that in NTE the components of the dynamical matrix have smaller values with decreasing volume. Finally, an alternative way to express $\alpha_{V}$ is to make use of thermodynamical relations and to write its expression in terms of vibrational entropy\cite{0953-8984-17-4-R03.jour.phys.cond.mat..Barrera.Bruno.Barron.NTE-rew.}
\begin{equation}
\alpha_{V}=\frac{1}{B}\sum_{\vec{q},j}\frac{\partial{S_{j}\left(\vec{q}\right)}}{\partial{V}},
\label{eq:alphav-entrspy}
\end{equation}
From Eq. \ref{eq:alphav-entrspy} it can be seen that the vibrational entropy $S$ must increase with decreasing volume for NTE to occur.

\subsection{Correlation functions}
\label{corr.funct.}
Mean square displacements (MSD) are a special case of the displacement time correlation functions derived by for example Maradudin.\cite{maradudin1971-harm-appr} Equal time displacement correlation function can be written as
\begin{eqnarray}
\left\langle u_{\alpha}\left(l\kappa\right)u_{\beta}\left(l'\kappa'\right)\right\rangle=\frac{\hbar}{2N\left(M_{\kappa}M_{\kappa'}\right)^{1/2}}
\nonumber\\
\times{}\sum_{\vec{q},j}\frac{e_{\alpha}\left(\kappa|\vec{q}j\right)e^{*}_{\beta}\left(\kappa'|\vec{q}j\right)}{\omega_{j}\left(\vec{q}\right)}
\nonumber\\
\times{}e^{i\vec{q}\cdot\left[\vec{x}\left(l\right)-\vec{x}\left(l'\right)\right]}\coth\left[\frac{\hbar\omega_{j}\left(\vec{q}\right)}{2k_{B}T}\right],
\label{eq:equaltcorr}
\end{eqnarray}
where $N$ is the number of unit cells (and $\vec{q}$-points) and $k_{B}$ is the Boltzmann constant. In the special case of $l=l'$ and $\kappa=\kappa'$, Eq. \ref{eq:equaltcorr} is called the equal time displacement autocorrelation function, vibrational amplitude, or MSD. From now on, all discussed displacement correlation functions are equal time displacement correlation functions and corresponding autocorrelation functions, if not otherwise mentioned. To analyze the NTE phenomenon, it is useful to know how the atoms move relative to each other. Mean square relative displacement (MSRD) in the interatomic direction between atoms $\left(l\kappa\right)$ and $\left(l'\kappa'\right)$, also known as the parallel MSRD, is defined as\cite{PhysRevB.14.1514.Beni.EXAFS1976}
\begin{equation}
\left\langle u^{2}_{\parallel}\right\rangle=\left\langle \left[\left(\vec{u}\left(l\kappa\right)-\vec{u}\left(l'\kappa'\right)\right)\cdot\vec{\hat{r}}\right]^{2}\right\rangle,
\label{eq:parallelMSD}
\end{equation}   
where $\vec{\hat{r}}$ is unit vector in the direction of vector between atoms $\left(l\kappa\right)$ and $\left(l'\kappa'\right)$. Only the difference of $l$ and $l'$ matters, not their absolute values, thus $l=0$ is chosen. Writing out the component form of Eq. \ref{eq:parallelMSD}, the parallel MSRD is
\begin{eqnarray}
\left\langle u^{2}_{\parallel}\right\rangle=\sum_{\alpha,\beta}r_{\alpha}r_{\beta}\left.[\left\langle u_{\alpha}\left(0\kappa\right)u_{\beta}\left(0\kappa\right)\right\rangle \right.
\nonumber\\
\left.+\left\langle u_{\alpha}\left(l'\kappa'\right)u_{\beta}\left(l'\kappa'\right)\right\rangle-2\left\langle u_{\alpha}\left(0\kappa\right)u_{\beta}\left(l'\kappa'\right)\right\rangle\right.].
\label{eq:parallelMSD2}
\end{eqnarray}  
Perpendicular MSRD is defined as\cite{0953-8984-13-34-324-Fornasini2001}
\begin{equation}
\left\langle u^{2}_{\perp}\right\rangle=\left\langle u^{2}_{R}\right\rangle-\left\langle u^{2}_{\parallel}\right\rangle,
\label{eq:pendendicularMSD}
\end{equation}
where
\begin{equation}
\left\langle u^{2}_{R}\right\rangle=\left\langle \left\|\vec{u}\left(0\kappa\right)-\vec{u}\left(l'\kappa'\right)\right\|^{2}\right\rangle,
\label{eq:absreldisp}
\end{equation}
which is the total MSRD of atoms $\left(l\kappa\right)$ and $\left(l'\kappa'\right)$. Writing Eq. \ref{eq:absreldisp} in the component form and substituting to Eq. \ref{eq:pendendicularMSD}, perpendicular MSRD can be written as
\begin{eqnarray}
\left\langle u^{2}_{\perp}\right\rangle=\sum_{\alpha,\beta}\left(\delta_{\alpha\beta}-r_{\alpha}r_{\beta}\right)\left.[\left\langle u_{\alpha}\left(0\kappa\right)u_{\beta}\left(0\kappa\right)\right\rangle \right.
\nonumber\\
\left.+\left\langle u_{\alpha}\left(l'\kappa'\right)u_{\beta}\left(l'\kappa'\right)\right\rangle-2\left\langle u_{\alpha}\left(0\kappa\right)u_{\beta}\left(l'\kappa'\right)\right\rangle\right.].
\label{eq:pendendicularMSD2}
\end{eqnarray}
Similar calculations involving the displacement correlation functions have been previously carried out by Nielsen and Weber, who used empirical interatomic potentials to investigate the correlation functions for \textit{d}-Si.\cite{0022-3719-13-13-005.Nielsen.Weber.displ.corr.1980,0305-4608-17-1-009.Weber.displ.corr.1987} Finally, the total MSD of atom $\left(l\kappa\right)$ is the sum of the autocorrelation functions along the different cartesian components
\begin{equation}
\left\langle u^{2}\right\rangle=\left\langle \left\|\vec{u}\left(l\kappa\right)\right\|^{2}\right\rangle=\sum_{\alpha}\left\langle u_{\alpha}\left(l\kappa\right)u_{\alpha}\left(l\kappa\right)\right\rangle.
\label{eq:absdisp}
\end{equation} 
In a cubic crystal, the total MSD (within numerical accuracy) is due to symmetry reasons equal to $3\left\langle u_{\alpha}\left(l\kappa\right)u_{\alpha}\left(l\kappa\right)\right\rangle$, for any $\alpha=1,2,3$.

Because the preceding displacement correlation functions are valid only within harmonic approximation, they cannot be used to describe the thermal expansion phenomenon as such. If QHA is assumed to be valid, the relation (Eq. \ref{eq:gruneisenparameter}) for the volume dependence of the phonon frequencies can be used. Inspection of Eq. \ref{eq:equaltcorr} shows that the two terms including the hyperbolic cotangent and the inverse of the frequency are always positive. As in Sec. \ref{L.dyn.therm.exp.}, from Eq. \ref{eq:CTEV2} and Eq. \ref{eq:gruneisenparameter} it can be seen that in NTE, the change in the frequencies must be in the same direction as the change in the volume. In general, the change of Eq. \ref{eq:equaltcorr} with respect to volume also depends on the behaviour of the phonon eigenvector and its components $e_{\alpha}\left(\kappa|\vec{q}j\right)$ as a function of volume. We show first with heuristic deduction that the volume dependence of Eq. \ref{eq:absdisp} in the case of \textit{d}-Si is only due to the volume dependence of the frequencies. Eq. \ref{eq:absdisp} reads
\begin{equation}
\left\langle u^{2}\right\rangle=C\sum_{\alpha}\sum_{\vec{q},j}\left|e_{\alpha}\left(\kappa|\vec{q}j\right)\right|^{2}\xi\left[\omega_{j}\left(\vec{q}\right),T\right],
\end{equation}
where
\begin{equation}
C=\frac{\hbar}{2NM_{\kappa}},
\end{equation}
and
\begin{equation}
\xi\left[\omega_{j}\left(\vec{q}\right),T\right]=\omega_{j}\left(\vec{q}\right)^{-1}\coth\left[\frac{\hbar\omega_{j}\left(\vec{q}\right)}{2k_{B}T}\right].
\end{equation}
The frequencies are not dependent on the cartesian index $\alpha$, and $\xi\left(\omega_{j}\left(\vec{q}\right)\right)$ can be moved out of the sum over $\alpha$
\begin{equation}
\left\langle u^{2}\right\rangle=C\sum_{\vec{q},j}\xi\left[\omega_{j}\left(\vec{q}\right),T\right]\sum_{\alpha}\left|e_{\alpha}\left(\kappa|\vec{q}j\right)\right|^{2}.
\label{eq:absdisp2}
\end{equation}
From Eq. \ref{eq:orthonorm} it follows that the sum over $\alpha$ in Eq. \ref{eq:absdisp2} with any chosen $j$ and $\vec{q}$ must have value $1/2$, in the case of \textit{d}-Si with two equivalent atoms in the primitive unit cell. Eq. \ref{eq:absdisp} for \textit{d}-Si is then
\begin{equation}
\left\langle u^{2}\right\rangle_{\textit{d}-Si}=\frac{C}{2}\sum_{\vec{q},j}\xi\left[\omega_{j}\left(\vec{q}\right),T\right].
\label{eq:absdispdiamond}
\end{equation}
Preceding is also true for structures where the lattice constants have been displaced in a way that preserves the original symmetry and it is also confirmed for \textit{d}-Si by means of a direct calculation. This implies that in NTE, the MSD of a particular atom $\left(l\kappa\right)$ and a particular mode $j$ increases with decreasing volume. The reason for this behaviour becomes clearer when the hyperbolic cotangent is written in terms of the Bose-Einstein distribution function $\overline{n}_{\vec{q}j}$. Using the definition of the hyperbolic cotangent, the last term in Eq. \ref{eq:absdispdiamond} is
\begin{equation}
\xi\left[\omega_{j}\left(\vec{q}\right),T\right]=\omega_{j}\left(\vec{q}\right)^{-1}\left(2\overline{n}_{\vec{q},j}+1\right),
\end{equation}
In NTE, Eq. \ref{eq:alphav-entrspy} and Eq. \ref{eq:absdispdiamond} actually have their values because of the same underlying reason. In the following, the similarity between these quantities, MSD for \textit{d}-Si and $S$, in NTE is considered when the temperature increases. If NTE occurs, the phonon frequency of the state $\left(\vec{q}j\right)$ contributing to NTE must have lower values when the volume decreases. The same effect occurs, when the considered phonon mode corresponds to a positive CTE and the volume increases. For the phonon modes corresponding to NTE, the preceding implies that there are more lower frequency states for phonons to occupy ($\overline{n}_{\vec{q}j}$ have higher values for lower $\omega_{j}\left(\vec{q}\right)$ in a particular temperature $T$). Therefore, the entropy increases and according to Eq. \ref{eq:alphav-entrspy}, NTE occurs when the opposite effect of the phonon modes corresponding to positive CTE is weaker. Similar conclusion applies to Eq. \ref{eq:absdispdiamond} in NTE. If the state $\left(\vec{q}j\right)$ has lower values of frequency when the volume decreases, there are more low frequency states that the phonons can occupy and $\xi\left[\omega_{j}\left(\vec{q}\right),T\right]$ has larger values, resulting in larger MSD values for the phonon modes contributing to NTE. The opposite is true for the phonon modes contributing to positive CTE. For both negative and positive CTE, the total MSD increases with increasing temperature. As a summary, in NTE, the MSD and entropy of the phonon modes corresponding to positive CTE increase less than the MSD and entropy of the phonon modes corresponding to NTE.

To study the effect of thermal expansion phenomenon to the MSD as a function of temperature, we define the following dimensionless quantity
\begin{equation}
\left\langle u^{2}\right\rangle_{r}\equiv \frac{1}{N_{a}} \left(\frac{V_{0}}{\left\langle u^{2}\right\rangle_{V_{0}}}\frac{\partial\left\langle u^{2}\right\rangle}{\partial V}- \left\langle u^{2}\right\rangle_{r,T_{0}} \right),
\label{eq:deltatotalMSD1}
\end{equation}
which includes the volume derivatives of the total MSD, and where
\begin{equation}
\left\langle u^{2}\right\rangle_{r,T_{0}} = \frac{V_{0}}{\left\langle u^{2}\right\rangle_{V_{0},T_{0}}}\frac{\partial\left\langle u^{2}\right\rangle_{T_{0}}}{\partial V}.
\end{equation}
In Eq. \ref{eq:deltatotalMSD1}, $N_{a}$ is the number of atoms within the primitive unit cell, $V_{0}$ is the volume of the equilibrium structure, $\left\langle u^{2}\right\rangle_{V_{0}}$ is the total MSD for the equilibrium structure, and $T_{0}$ indicates the value of the function at $T=0$K. Eq. \ref{eq:deltatotalMSD1} measures the product of the total MSD volume derivative and the ratio of the equilibrium volume and the equilibrium values of the total MSD. The values of Eq. \ref{eq:deltatotalMSD1} are therefore negative for modes for which the values of the frequency decrease as the volume decreases, at a particular temperature $T$. This is based on the discussions above and the fact that the MSD (Eq. \ref{eq:absdisp}) have only positive values for all values of $T$. The values of the first term within the brackets in Eq. \ref{eq:deltatotalMSD1} differ from zero when $T \rightarrow 0$ K. For clarity, a normalization factor is defined (the second term within the brackets in Eq. \ref{eq:deltatotalMSD1}), which has a constant value for any chosen state $\left(\vec{q}j\right)$. The normalization factor ensures that Eq. \ref{eq:deltatotalMSD1} is equal to zero when $T \rightarrow 0$ K. The preceding normalization works properly since the first term within the brackets in Eq. \ref{eq:deltatotalMSD1} increases (positive CTE) or decreases (negative CTE) with all relevant $T>0$. In addition, a second normalizing factor $N_{a}$ is introduced to facilitate comparisons between structures with different-sized primitive unit cells.  $N_{a}$ is related to the number of phonon modes $N_{j}$ by the relation $N_{j}=3N_{a}$. The same scheme is implemented for the parallel and perpendicular MSRD to study changes in the relative displacements with respect to volume and their role in the NTE phenomenon. They are defined as
\begin{equation}
\left\langle u^{2}_{\parallel}\right\rangle_{r}\equiv \frac{1}{N_{a}}\left(\frac{V_{0}}{\left\langle u^{2}_{\parallel}\right\rangle_{V_{0}}}\frac{\partial\left\langle u^{2}_{\parallel}\right\rangle}{\partial V}-\left\langle u^{2}_{\parallel}\right\rangle_{r,T_{0}}\right),
\label{eq:deltatotalparal}
\end{equation} 
and   
\begin{equation}
\left\langle u^{2}_{\perp}\right\rangle_{r}\equiv \frac{1}{N_{a}}\left(\frac{V_{0}}{\left\langle u^{2}_{\perp}\right\rangle_{V_{0}}}\frac{\partial\left\langle u^{2}_{\perp}\right\rangle}{\partial V}-\left\langle u^{2}_{\perp}\right\rangle_{r,T_{0}}\right).
\label{eq:deltatotalMSD1perdendic}
\end{equation}    
The preceding quantities in Eqs. \ref{eq:deltatotalMSD1} and \ref{eq:deltatotalparal}--\ref{eq:deltatotalMSD1perdendic} are named as relative MSD and MSRD volume derivatives, and these quantities are denoted as MSD-VD and MSRD-VD, respectively.

Properties analogous to the MSD-VD and MSRD-VD introduced above can be derived also for equal time momentum correlation functions, which can be written within harmonic approximation as\cite{maradudin1971-harm-appr}
\begin{eqnarray}
\left\langle p_{\alpha}\left(l\kappa\right)p_{\beta}\left(l'\kappa'\right)\right\rangle=\frac{\hbar}{2N}\left(M_{\kappa}M_{\kappa'}\right)^{1/2}
\nonumber\\
\times{}\sum_{\vec{q},j}e_{\alpha}\left(\kappa|\vec{q}j\right)e^{*}_{\beta}\left(\kappa'|\vec{q}j\right)\omega_{j}\left(\vec{q}\right)
\nonumber\\
\times{}e^{i\vec{q}\cdot\left[\vec{x}\left(l\right)-\vec{x}\left(l'\right)\right]}\coth\left[\frac{\hbar\omega_{j}\left(\vec{q}\right)}{2k_{B}T}\right]=
\nonumber\\
M_{\kappa}M_{\kappa'} \omega^{2}_{j}\left(\vec{q}\right) \left\langle u_{\alpha}\left(l\kappa\right)u_{\beta}\left(l'\kappa'\right)\right\rangle,
\label{eq:equaltcorrmomentum}
\end{eqnarray}
where $p_{\alpha}\left(l\kappa\right)$ is the momentum for the atom $\left(l\kappa\right)$ in the direction $\alpha$. The difference is that the change of the mean square momentum (MSM) with respect to the volume, derived from Eq. \ref{eq:equaltcorrmomentum}, is opposite in direction to the MSD-VD and MSRD-VD. This means that if the volume decreases and the temperature increases for example by 1 K (NTE occurs), the phonon modes contributing to NTE (positive CTE) resist (assist) the increase of the MSM, but since the changes in the frequencies are small ( $\left|\Delta \omega_{j}\left(\vec{q}\right)\right|/ \omega_{j}\left(\vec{q}\right) << 1$), MSM increases in both cases. 

Finally, few remarks about the method of calculating CTE from Eq. \ref{eq:CTEV2} or \ref{eq:alphav-entrspy} and the related properties calculated from Eq. \ref{eq:deltatotalMSD1} are necessary. In general, the Gr\"{u}neisen parameters are not independent of the temperature, which also means that $\omega_{j}\left[\vec{q},V\left(T\right),T\right]$. When the temperature independence is assumed it follows from Eq. \ref{eq:CTEV2} and \ref{eq:alphav-entrspy} that the phonon modes corresponding to positive CTE (NTE) have positive (negative) contribution to CTE at all finite $T>0$. The same features result for the MSD-VD and MSRD-VD and the related quantities derived for momentum. 

\subsection{Studied structures and computational details}
\label{Struct.calc.param.}
\begin{figure}
\includegraphics[width=0.48\textwidth]{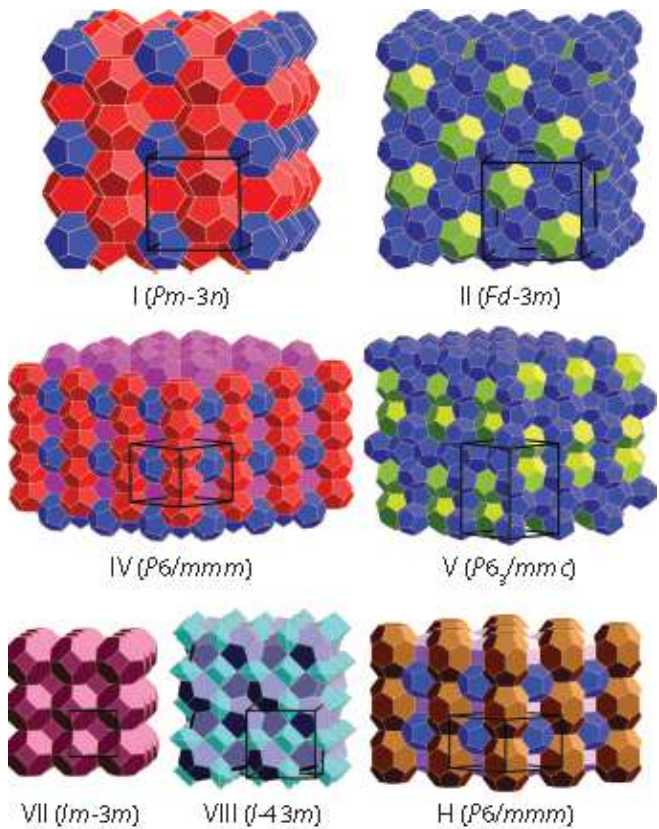}
\caption{Schematic figures of the silicon clathrate structures included in this work. The vertices of the polyhedral cages present silicon atoms. Crystallographic unit cell edges are drawn in black. For a more comprehensive description of the structural characteristics, see.\cite{karttunen2010structuralprinc} (Color online)} 
\label{fig:structures}
\end{figure}

The structural characteristics of the studied seven silicon clathrate frameworks are illustrated in Fig. \ref{fig:structures} and described in Table \ref{tab:parameters}.
\begin{table*}
\caption{Structural data and computational details for the studied structures.}
		\begin{tabular}{ccccccc}
		\hline\hline
		Structure      & Space Group                  & Atoms/cell\footnote[1]{The number of atoms in the primitive cell.} & Elect. (k1,k2,k3)\footnote[2]{The mesh used for the electronic $\vec{k}$-sampling.} &  Phon. (q1,q2,q3)\footnote[3]{The meshes used for phonon calculations, PDOS calculations, and correlation function calculations, respectively. The meshes for the PDOS and correlation function calculations were Fourier interpolated from the meshes used in the phonon calculations.} & PDOS (q1,q2,q3)$^c$ & Corr. (q1,q2,q3)$^c$ \\\hline
	  \textit{d}-Si  & $Fd\bar{3}m\left(227\right)$ & 2  &  16,16,16   & 8,8,8 & 80,80,80 & 20,20,20 \\ 
	  I              & $Pm\bar{3}n\left(223\right)$ & 46 &   4,4,4     & 4,4,4 & 30,30,30 & -        \\
		II             & $Fd\bar{3}m\left(227\right)$ & 34 &   4,4,4     & 4,4,4 & 30,30,30 & 8,8,8    \\
		IV             & $P6/mmm\left(191\right)$     & 40 &   4,4,4     & 3,3,3 & 30,30,30 & -        \\
		V              & $P6_{3}/mmc\left(194\right)$ & 68 &   4,4,3     & 4,4,3 & 30,30,30 & -        \\
		VII            & $Im\bar{3}m\left(229\right)$ & 6  &   8,8,8     & 8,8,8 & 60,60,60 & -        \\
		VIII           & $I\bar{4}3m\left(217\right)$ & 23 &   6,6,6     & 4,4,4 & 40,40,40 & -        \\
		H              & $P6/mmm\left(191\right)$     & 34 &   4,4,4     & 3,3,3 & 30,30,30 & -        \\
		\hline\hline
		\end{tabular}
		\label{tab:parameters}
\end{table*}
The semiconducting clathrate structures are classified according to the polyhedral cages they are composed of. The structural properties of the clathrates have been reviewed in e.g. \cite{Rogl2006,0036-021X-73-9-R06.Kovnir.Shevelkov.clath.review.2004.,Shevelkov-2011-zintl-clath}

The \textit{ab initio} density functional calculations to optimize the structures and to calculate the phonon dispersion relations were carried out with the Quantum Espresso program package (QE, version 5.0.3).\cite{QE-2009} The Si atoms were described by Martins-Troullier-type norm-conserving pseudopotentials \cite{PhysRevB.43.1993.Troullier.Martins.pz.n.nc.pseudot.1991,pseudopotentials} and the Local Density Approximation (LDA) was used as the exchange-correlation energy functional.\cite{PhysRevB.23.5048-Perdew-Zunger-LDA} In all calculations, the following kinetic energy cutoffs have been used: 40 Ry for wavefunctions and 160 Ry for charge densities and potentials. The applied $\vec{k}$- and $\vec{q}$-sampling for each studied structure is listed in Table \ref{tab:parameters}. The $\vec{q}$-meshes for the phonon density of states (PDOS) and correlation function calculations were Fourier interpolated from the mesh used in the corresponding phonon calculation (QE module matdyn.x).\cite{RevModPhys.73.515-Baroni-2001-DFTP} We carried out convergence tests for both SCF and phonon frequency calculations with different $\vec{k}$-meshes. The lattice constant for the optimized structure was practically identical for (8,8,8) and (16,16,16) $\vec{k}$-meshes and the final energy was about 5*10$^{-3}$ eV lower with the (16,16,16) mesh. The maximum differences between the phonon frequencies obtained with the (8,8,8) and (16,16,16) $\vec{k}$-meshes are about 1.5\%. Since a reasonable convergence was found for the (8,8,8) $\vec{k}$-mesh in the case of \textit{d}-Si, the $\vec{k}$-meshes listed in in Table \ref{tab:parameters} for the clathrate frameworks were chosen as a compromise between accuracy and computational cost. For \textit{d}-Si, the use of the denser (16,16,16) $\vec{k}$-mesh was still computationally feasible. We have also confirmed the convergence of the calculated phonon frequencies with respect to the applied $\vec{q}$-meshes. Both the lattice constants and the atomic positions of the studied structures were fully optimized (applying the space group symmetries listed in Table \ref{tab:parameters}). The Gr\"{u}neisen parameters were calculated by displacing the structures from the equilibrium lattice constant by $\pm$0.5\%, optimizing the atomic positions, and calculating the numerical derivatives using central differences. Bulk moduli were calculated by displacing the structures from the equilibrium lattice constant up to $\pm$2\% with a step size of $\pm$0.5\%  and calculating the bulk modulus using the Murnaghan's equation of state (QE module ev.x.). Phonon density of states values were calculated using the tetrahedron method (QE module matdyn.x).\cite{PhysRevB.49.16223-Blochl-tetrahedron} In all cases where a summation over $\vec{q}$ is involved, the appropriate weights to normalize the sums have been used. The algorithms for calculating the thermodynamic quantities and the correlation functions were implemented as Matlab routines.

Calculation of the correlation functions involves some numerical issues, which need to be discussed. To our knowledge, there is no way to label the phonon modes uniquely in a point of degeneracy when the diagonalization of the dynamical matrix in Eq. \ref{eq:eigenvalueeq} is done numerically. This is true also for the QE program package and probably results in minor numerical inaccuracies in the calculation of the correlation functions. Some phonon labeling could be carried out by continuity arguments for the phonon eigenvectors in particular direction when using relatively dense $\vec{q}$-meshes, but in the absence of a rigorous general approach, the labeling of the phonon modes is left to the default algorithm in QE. The numerical accuracy is also sensitive to the applied Fast Fourier Transform (FFT) mesh. In this work, the default settings of QE were used to choose the FFT grid.

\section{Results and discussion}
\label{Results and discussion}
\subsection{Optimized structures and phonon dispersion relations}
\label{Opt.struct.disp.rel.}
Table \ref{tab:calcresults} lists the lattice constants, the relative energies $\Delta E$ with respect to \textit{d}-Si, and the bulk moduli for the optimized silicon clathrate frameworks.
\begin{table}
\caption{Lattice constants $a,c$, relative energies ($\Delta E$), and bulk moduli ($B$) for the optimized structures.}
		\begin{tabular}{ccccc}
		\hline\hline
			  Structure & $a$(\AA) & $c$(\AA) &  $\Delta E$ (eV/atom) & $B$ (GPa)   \\\hline
 \textit{d}-Si    & 5.38   &    $-$     &    0.00   &   96.5     \\
	           I    & 10.08  &    $-$     &    0.09   &   84.1     \\
	          II    & 14.52  &    $-$     &    0.08   &   82.4     \\
				    IV    & 10.07  &    10.23   &    0.11   &   78.5     \\
				     V    & 10.26  &    16.80   &    0.08   &   82.1     \\
				   VII    & 6.63   &    $-$     &    0.27   &   72.6     \\
				  VIII    & 9.95   &    $-$     &    0.10   &   86.6     \\
				  H       & 10.36  &    8.43    &    0.11   &   79.6     \\
		\hline\hline
		\end{tabular}
		\label{tab:calcresults}
\end{table}
As could be expected, the structure II, which has been synthetized as a nearly guest-free silicon framework, shows the lowest relative energy among the studied clathrate frameworks. It is followed closely by structure V, which is a hypothetical hexagonal modification of the structure II.\cite{karttunen2010structuralprinc} The predicted relative energies are in good agreement with previous results obtained with the PBE0 hybrid density functional method and localized Gaussian type basis sets (GTO).\cite{karttunen2010structuralprinc} The structure VII shows the largest difference in $\Delta E$ in comparison to the PBE0 results (0.03 eV/atom), while for the other structures the difference is only 0.01 eV/atom. The structure VII is a rather strained one with a high relative energy and it is not experimentally as relevant as the other studied structures. The predicted bulk moduli are systematically somewhat smaller in comparison to previous results obtained with the PBE0 functional,\cite{doi:10.1021/jp205676p.oma.Antti.mech.prop.} but the result for \textit{d}-Si is similar to previous LDA-predictions.\cite{PhysRevB.32.3792.Nielsen.Stresses.in.semic.1985} The previous LDA predictions for the clathrate frameworks I and II (87 and 87.5 GPa, respectively) are also comparable to the present results.\cite{San-Miguel.et.al.2002.clath-PhysRevB.65.054109} LDA slightly underestimates the bulk modulus ($\approx$1.3\% in the case of \textit{d}-Si) in comparison to experiment, the experimental value for \textit{d}-Si being 97.8 GPa.\cite{PhysRev.161.756.Hall.Bulk.modulus.Si.experim.1967} The difference between calculated and experimental lattice constant value is approximapdftely 1\%.\cite{PhysRevLett.72.3133-Basile-et.al.Si-lat.const.} The experimental value of the bulk modulus for the silicon clathrate framework II has been found to be $90 \pm 5$ GPa.\cite{San-Miguel.et.al.1999.clath-PhysRevLett.83.5290}

\begin{figure}
\includegraphics[width=0.47\textwidth]{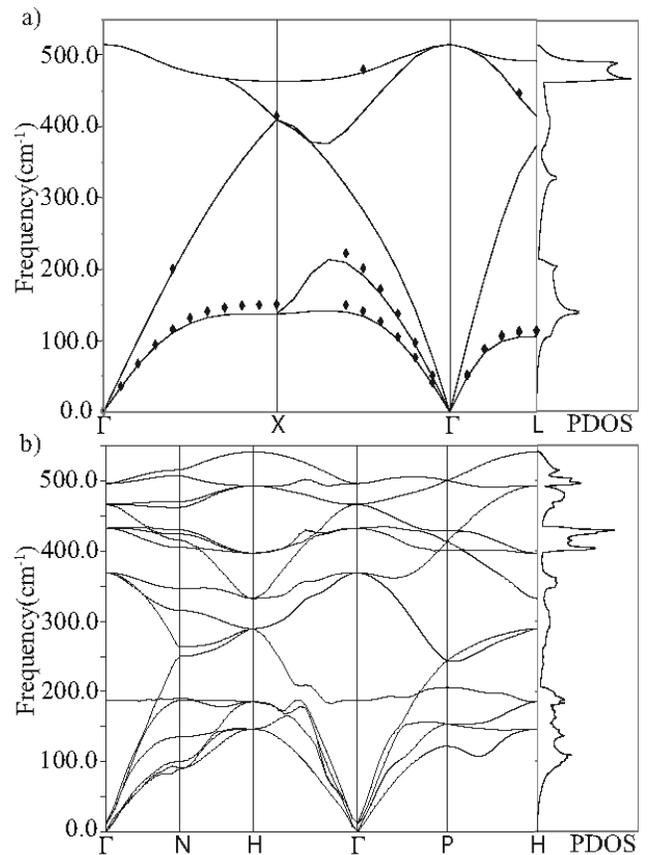}
\caption{Phonon dispersion relations along high symmetry paths in the first Brillouin zone for a$)$ \textit{d}-Si (experimental values indicated by diamonds\cite{PhysRevB.6.3777.Nilsson.dispersion.experim.Sialpha}) and b$)$ for structure VII.}
\label{fig:disp_alpha-VII}
\end{figure}

\begin{figure*}
\includegraphics[width=0.99\textwidth]{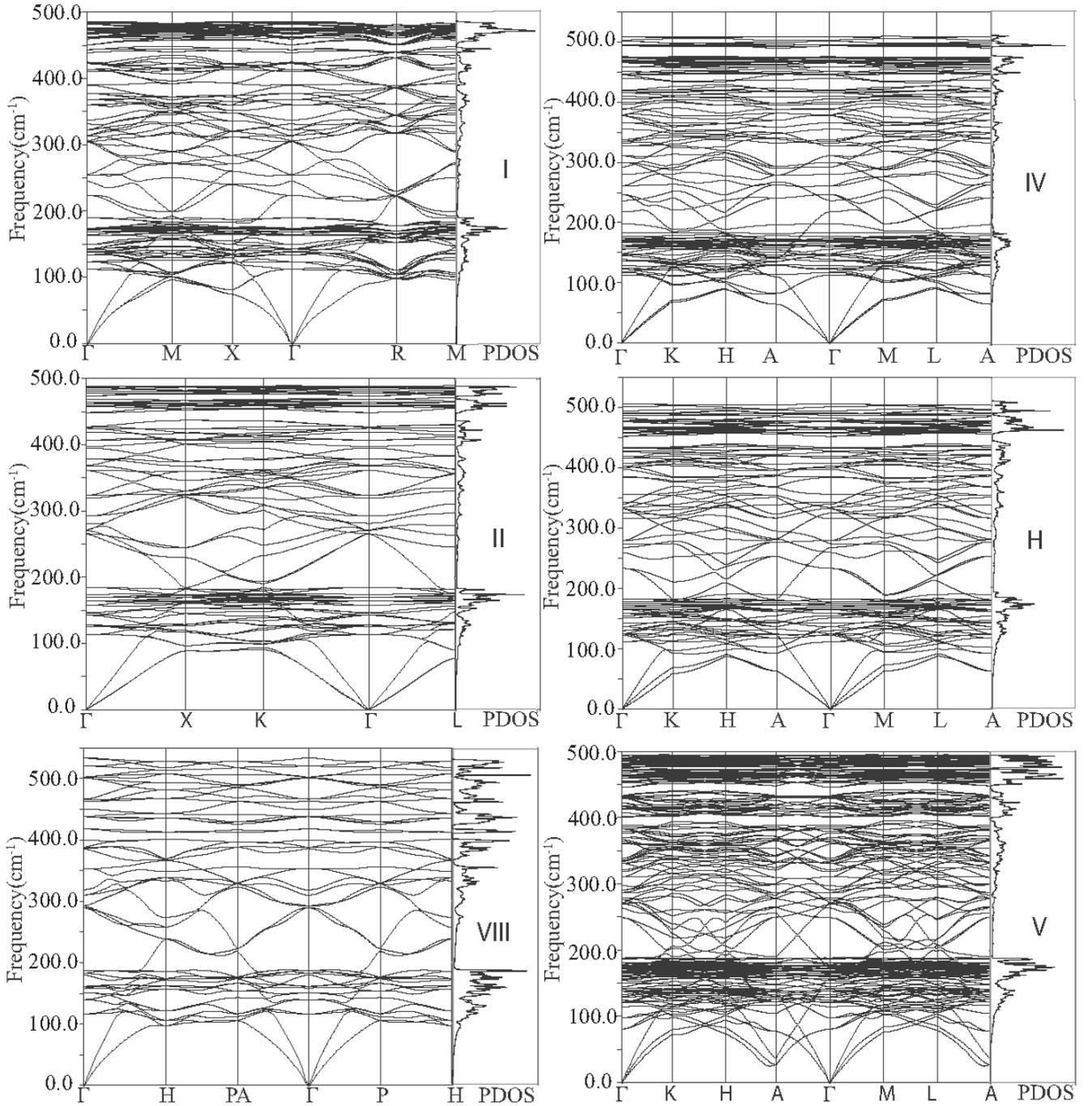}
\caption{Phonon dispersion relations along high symmetry paths in the first Brillouin zone for structures I, II, IV, V, VIII, and H.}
\label{fig:disp-kaikki}
\end{figure*}

Figures \ref{fig:disp_alpha-VII} and \ref{fig:disp-kaikki} illustrate the phonon dispersion relations in the studied structures and the corresponding phonon density of states (PDOS). 
Fig. \ref{fig:disp_alpha-VII} also shows the experimental phonon dispersion data for \textit{d}-Si.\cite{PhysRevB.6.3777.Nilsson.dispersion.experim.Sialpha} The LDA results for \textit{d}-Si are in relatively good agreement with the experiment, the differences being the largest for large wave vectors (the max. difference is approximately 7\%). The estimated experimental error is $\pm$1 cm$^{-1}$.\cite{PhysRevB.6.3777.Nilsson.dispersion.experim.Sialpha} The LDA phonon dispersion data for the clathrate framework II are also in agreement with previous computational results.\cite{PhysRevB.74.014109.Tang.et.al.thermal.prop.clathII} The PDOS of the silicon clathrate frameworks I and II have also been determined experimentally using inelastic neutron scattering for structures containing some potassium and sodium impurities (e.g. $\textrm{Na}:\textrm{Si}$ ratio 1:34). Despite the impurities, the experimental results show similar main features when compared to the results shown in Fig. \ref{fig:disp-kaikki}, such as the high PDOS values at the frequencies of approximately 175 cm$^{-1}$.\cite{Melinon-PDOS-clath-I-II.1999-PhysRevB.59.10099} In the structure VII (Fig. \ref{fig:disp_alpha-VII}b), a significant decrease of the lowest optical modes towards the acoustic modes can be observed, in particular for small wave vectors. This indicates strong anomalies in the interatomic forces. The other clathrate frameworks show very similar phonon dispersion relations compared to each other. The structures other than \textit{d}-Si and structure VII have a phonon band gap within the frequency range of 400--475 cm$^{-1}$, the structure V even showing two phonon band gaps at frequencies of about 440 and 390 cm$^{-1}$ (Fig. \ref{fig:disp-kaikki}).

\subsection{Thermal properties from the quasiharmonic approximation}
\label{Therm.prop.QHA}
\begin{figure}
\includegraphics[width=0.48\textwidth]{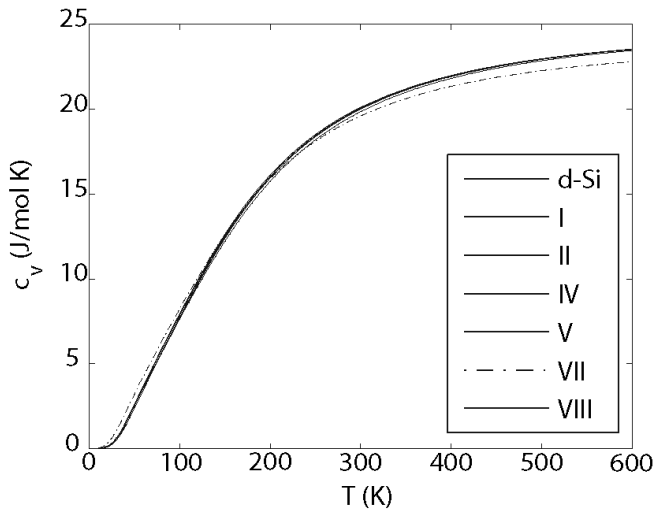}
\caption{Heat capacities at constant volume for studied structures. Only the structure VII is slightly different from the other structures (indicated with a dash-dotted line).}
\label{fig:heatcapac}
\end{figure}

\begin{figure}
\includegraphics[width=0.48\textwidth]{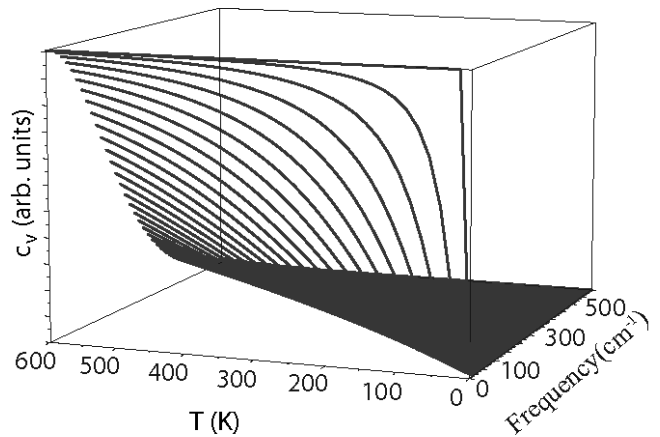}
\caption{Heat capacity at constant volume as a function of temperature and frequency in QHA. Every separate line represents the value of Eq. \ref{eq:heatcapacity} for a particular phonon frequency as a function of the temperature.}
\label{fig:heatcapactempdep}
\end{figure}

The calculated heat capacities at constant volume are shown in Fig. \ref{fig:heatcapac}.
The results are very similar for all structures, the strained structure VII showing the largest difference to the other structures. The results for the structure II are in agreement with previous experimental and computational results.\cite{Biswas.Nolas:033535.Therm.prop.ofSi136.2008} The reason for the deviation observed for the heat capacity of structure VII can be seen from Fig. \ref{fig:heatcapactempdep}, where Eq. \ref{eq:heatcapacity} has been plotted.
At low $T$, only the low frequency modes affect the heat capacity significantly. When $T$ increases, the contribution of the low frequency modes does not increase as rapidly as the contribution of the higher frequency modes. Since the structure VII has more low frequency modes and less higher frequency modes in comparison to the other structures, it has a slightly higher heat capacity at low temperature and lower heat capacity at higher temperatures. This also affects the CTE of the structure VII. The very similar heat capacities suggest that in the studied structures, and within QHA, the Gr\"{u}neisen parameters and the differences in the values of the bulk moduli have the greatest effect on the magnitude of CTE.

\begin{figure*}
\includegraphics[width=0.99\textwidth]{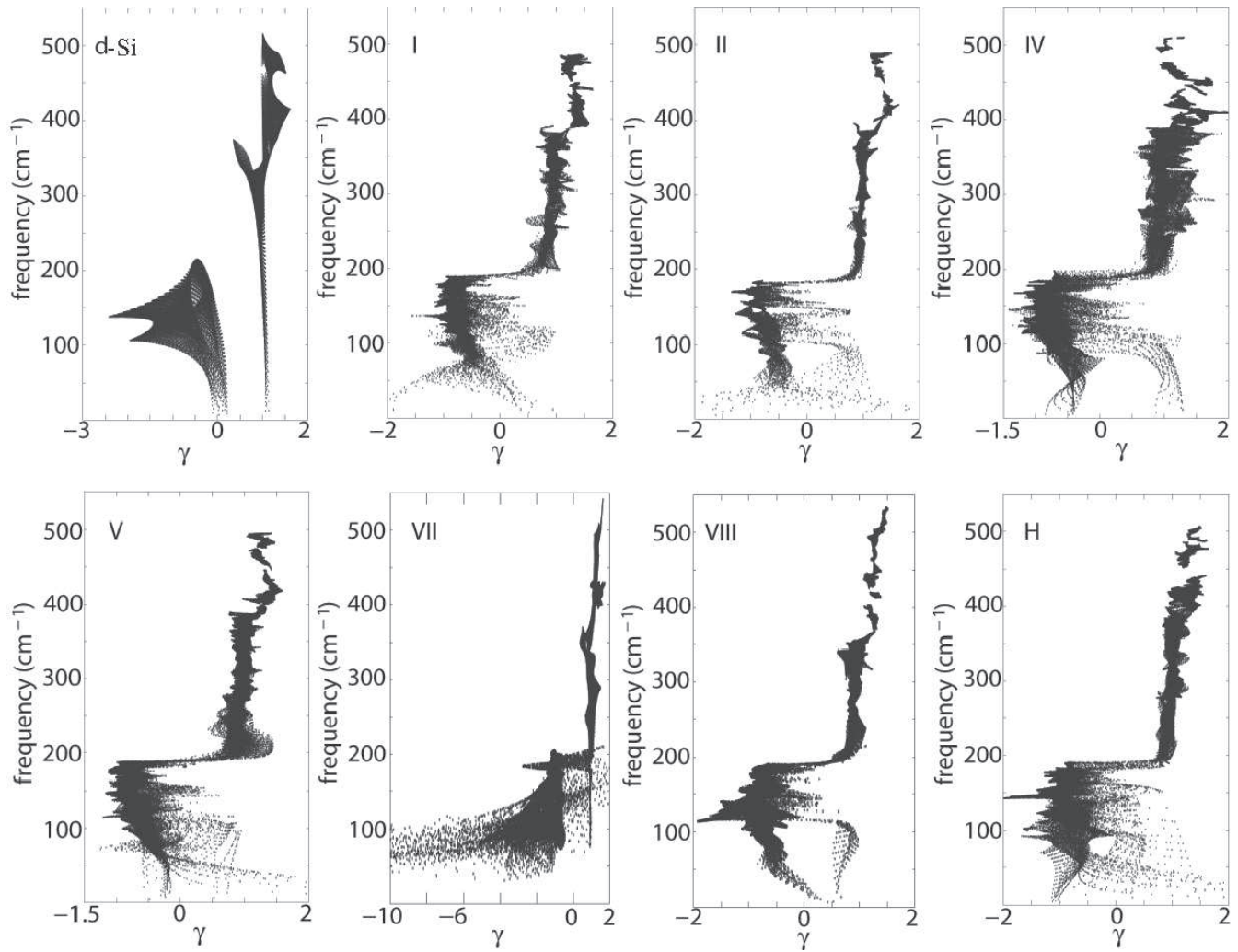}
\caption{Gr\"{u}neisen parameters for the studied structures, each dot represents one mode at a particular $\vec{q}$-point.}
\label{fig:gruneisenall}
\end{figure*}

Fig. \ref{fig:gruneisenall} shows the Gr\"{u}neisen parameters for the studied structures as a function of the phonon frequency.
For every structure, approximately 75000 frequency values have been used to generate the Gr\"{u}neisen parameter data. The Gr\"{u}neisen parameters predicted for \textit{d}-Si and the structure II are in agreement with previous computational results.\cite{PhysRevB.74.014109.Tang.et.al.thermal.prop.clathII} The structure VII has distinctly different Gr\"{u}neisen parameters from the other structures, showing negative values that are approximately 10 times larger than for the other structures. An interesting feature is that the Gr\"{u}neisen parameters only have negative values when the phonon frequency is under a certain threshold that is ~220 cm$^{-1}$ for \textit{d}-Si and VII and ~190 cm$^{-1}$ for the other structures. However, in every structure, the acoustic modes with frequencies significantly below 100 cm$^{-1}$ do have positive Gr\"{u}neisen parameter values at some $\vec{q}$. For the longitudal phonon mode of \textit{d}-Si, the Gr\"{u}neisen parameter values are positive for all wave vectors. On the other hand, the two transverse acoustic phonon modes have positive values of Gr\"{u}neisen parameter only at few wave vectors. The Gr\"{u}neisen parameters of the clathrate frameworks show features similar to amorphous silicon,\cite{PhysRevLett.79.1885-Allen-gruneisen-amorph.-Si} where the Gr\"{u}neisen parameter values are spread out at low frequency values. This is in contrast to \textit{d}-Si, where the Gr\"{u}neisen parameters show a rather symmetric shape as a function of frequency in comparison to the studied clathrate frameworks and amorphous silicon.

\begin{figure}
\includegraphics[width=0.48\textwidth]{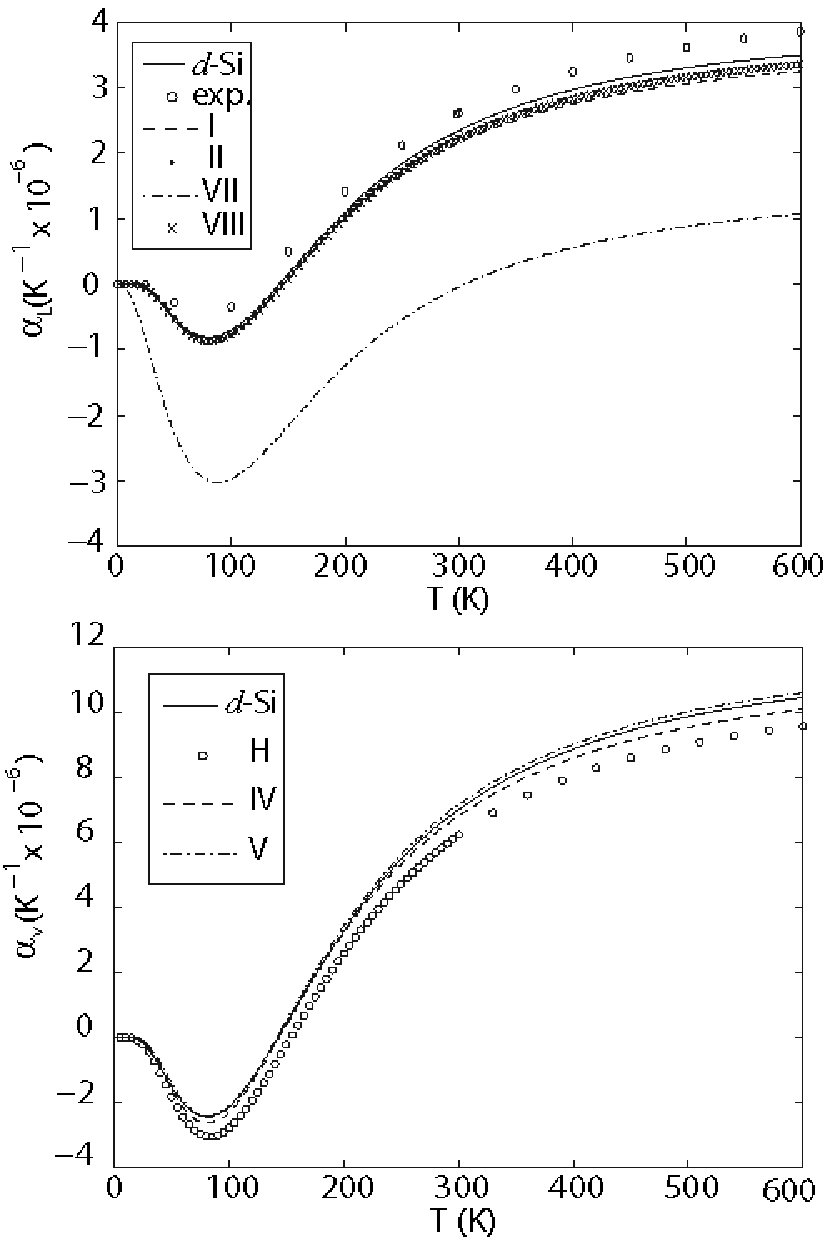}
\caption{Top: Linear CTE for the studied cubic structures. The experimental values for \textit{d}-Si are indicated with circles.\cite{Reeber1996259.Sialpha.thermalexp.experimental} Bottom: Volumetric CTE for the studied hexagonal structures, \textit{d}-Si is included as a reference.}
\label{fig:thermexp}
\end{figure}

The results in Figures \ref{fig:heatcapac} and \ref{fig:gruneisenall} suggest that only the structure VII should show significantly different CTE in comparison to the other studied structures. This is confirmed by the linear and volumetric CTE values shown in Fig. \ref{fig:thermexp} for all studied structures. 
For \textit{d}-Si, the experimental values for the linear CTE are also shown.\cite{Reeber1996259.Sialpha.thermalexp.experimental} The CTE values obtained from Eq. \ref{eq:CTEV2} and Eq. \ref{eq:alphav-entrspy} are identical. In comparison to experiment, LDA overestimates the magnitude of the NTE for \textit{d}-Si and underestimates the CTE at positive values. In all studied structures, excluding the  structure VII, the CTE values are very similar. The structure VII with distinctly different Gr\"{u}neisen parameters shows NTE behavior up to approximately 300 K and the NTE is nearly five times larger than for \textit{d}-Si. As discussed above, the structure VII has a very high relative energy and it is not experimentally as relevant as the other structures. However, it is still an interesting test case for detailed studies on the NTE phenomenon due to the small size of its primitive unit cell and its anomalous lattice dynamical properties.

\begin{figure}
\includegraphics[width=0.48\textwidth]{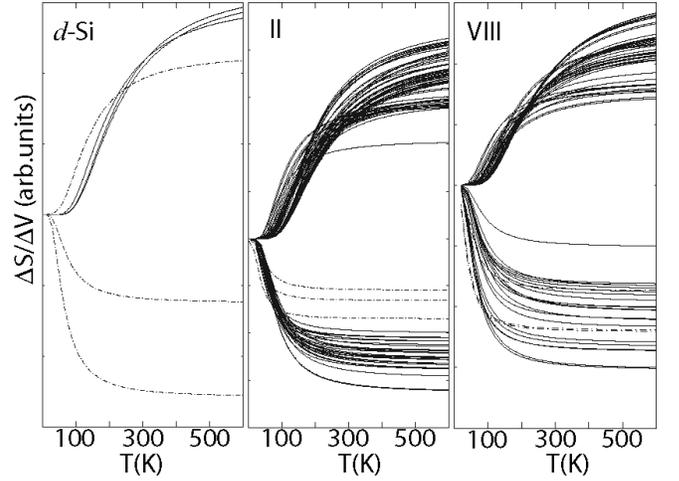}
\caption{$\Delta S/\Delta V$ for each phonon mode in the \textit{d}-Si, II, and VIII structures as a function of the temperature. The zero level is at the point where the lines representing the phonon modes coincide at $T=0$ K. The acoustic modes are indicated with dash-dotted lines.}
\label{fig:deltaS}
\end{figure}

Fig. \ref{fig:deltaS} illustrates the change in entropy $\Delta S$ with respect to the change in the volume $\Delta V$ for each phonon mode as a function of temperature for \textit{d}-Si, and the clathrate frameworks II and VIII.
In \textit{d}-Si, the longitudinal acoustic mode has a positive $\Delta S/\Delta V$ at all temperatures, having a positive contribution to CTE (Eq. \ref{eq:alphav-entrspy}). The transverse acoustic modes have a negative contribution to CTE, giving rise to the experimentally confirmed NTE behavior of \textit{d}-Si. In the structures II and VIII, according to QHA, all acoustic modes have a negative contribution to CTE at all considered temperatures. The reason for this was discussed at the end of Sec. \ref{corr.funct.}. An interesting feature of the structures II and VIII is that several optical modes (all optical modes in the structure II) have larger contributions to NTE than the acoustic modes when $T>$100 K. Similar crossing occurs between optical phonon modes having a positive contribution to CTE, as shown in Fig. \ref{fig:deltaS}. The difference is that the crossing occurs at $T \approx 200$K. For the studied structures, the maximum value of $\Delta S/\Delta V$, when the temperature changes by $\pm 1$K, is six orders of magnitude smaller than the absolute value of $S$ at the same temperature.

\subsection{Thermal expansion and displacement correlation functions}
\label{Therm.exp.cor.funct.}
\begin{figure}
\includegraphics[width=0.48\textwidth]{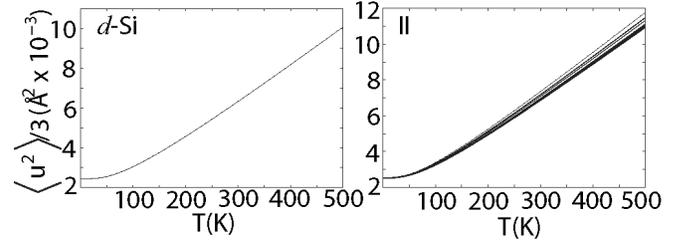}
\caption{Displacement autocorrelation functions for \textit{d}-Si and the clathrate framework II (Eq. \ref{eq:equaltcorr}). Each line represents the displacement autocorrelation function for one atom in the primitive unit cell.}
\label{fig:autocorr-alpha-II}
\end{figure}

\begin{figure}
\includegraphics[width=0.48\textwidth]{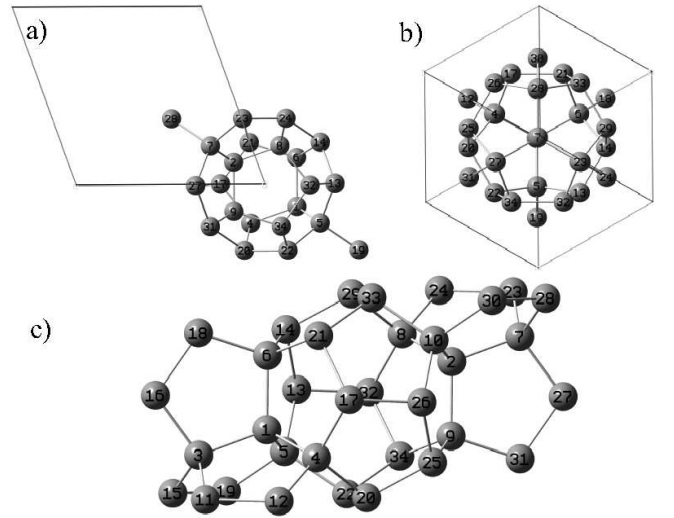}
\caption{The primitive unit cell of the clathrate framework II shown in a) [100] b) [111] and c) general direction. Unit cell edges are drawn in black.}
\label{fig:unit-cell-II}
\end{figure}

Fig. \ref{fig:autocorr-alpha-II} shows the displacement autocorrelation functions (Eq. \ref{eq:equaltcorr}) for all atoms within the primitive unit cells of \textit{d}-Si and clathrate II.
The results for \textit{d}-Si are in good agreement with previous results of Nielsen and Weber obtained using empirical potentials,\cite{0022-3719-13-13-005.Nielsen.Weber.displ.corr.1980} and with the results of Strauch \textit{et al.} obtained using \textit{ab initio} methods.\cite{Strauch-1996-436-ab-initio-MSD} The differences between \textit{d}-Si and the clathrate framework II are small. In the case of the structure II, the values are separated in three different groups corresponding to the three different Wyckoff positions in the crystal structure. The primitive unit cell of the structure II is shown in Fig. \ref{fig:unit-cell-II} to facilitate the discussion on the results related to the displacement correlation functions.
The highest MSD values correspond to atoms 1 and 2 in Fig. \ref{fig:unit-cell-II} (Wyckoff position 8\textit{a}), the intermediate values to atoms 3-10 (32\textit{e}), and the smallest values to atoms 11-34 (96\textit{g}). The MSD values calculated for the symmetry equivalent atoms in the structure II show a minor deviation from each other. The highest deviations correspond to atoms 11-34, when the max. difference within this group is approximately 1.8\%, which occurs at $T=500$ K. One possible reason for this is the issue of phonon mode labeling for degenerate modes during the diagonalization of the dynamical matrix, as discussed in the end of Sec. \ref{Struct.calc.param.}. The total MSD (Eq. \ref{eq:absdisp}) calculated for the structure VII is approximately 25\% higher than the total MSDs calculated for \textit{d}-Si and the structure II.

\begin{figure}
\includegraphics[width=0.48\textwidth]{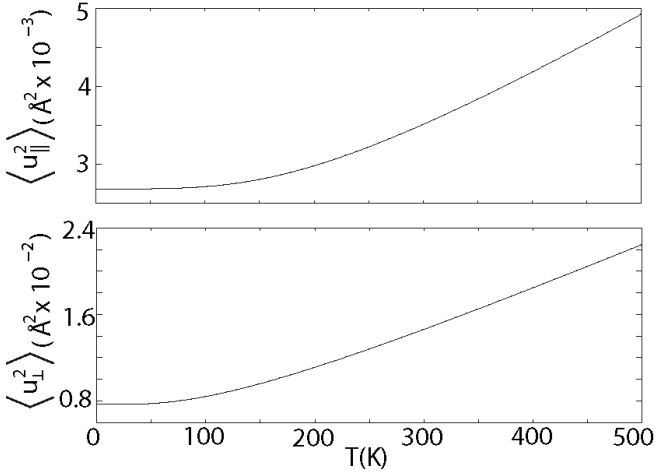}
\caption{Parallel and perpendicular correlation functions for \textit{d}-Si within same unit cell, i.e. for the nearest neigbours.}
\label{fig:si-alpha-para-pen-01_02}
\end{figure}

The parallel and perpendicular displacement correlation functions between the nearest neighbours in \textit{d}-Si ($l\kappa=l1;l\kappa'=l2$) are presented in Fig. \ref{fig:si-alpha-para-pen-01_02}.
\begin{figure}
\includegraphics[width=0.48\textwidth]{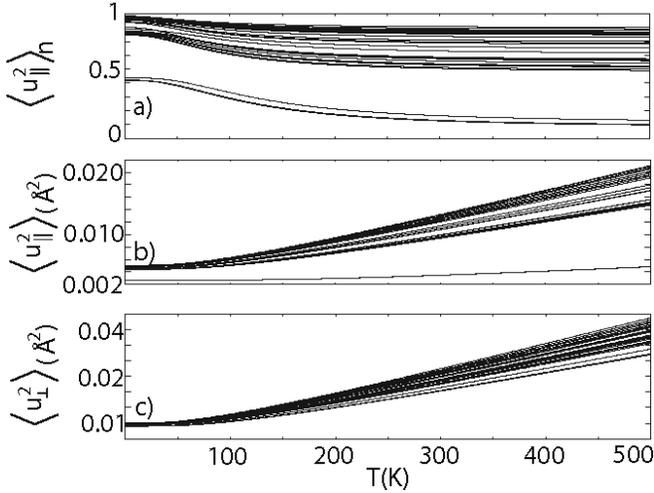}
\caption{Parallel and perpendicular MSRD within the same unit cell for the structure II. a) Normalized parallel MSRD. b) Parallel MSRD and c) perpendicular MSRD. Different lines represent different atoms $\kappa$, the second atom $\kappa'=2$ in all cases (see Fig. \ref{fig:unit-cell-II}).}
\label{fig:II-parallel-normalizedparal-pend-kp2}
\end{figure}
The perpendicular MSRDs have larger values than the parallel MSRDs. Similar behavior has been confirmed with experimental EXAFS studies and computational results for \textit{d}-Ge,\cite{PhysRevLett.82.4240-Dalba-1999-Ge-EXAFS}$^{,}$\cite{0022-3719-13-13-005.Nielsen.Weber.displ.corr.1980} and also for other materials such as CdTe with zincblende structure using EXAFS.\cite{0953-8984-24-11-115403.Abd.et.al.EXAFS.NTE.2012} In cubic crystal structures, the perpendicular MSRD values must be larger than the parallel MSRD values due to symmetry reasons. If a three-dimensional Cartesian coordinate system is considered to have one of its orthonormal basis vectors along the vector between the nearest neighbour atoms in \textit{d}-Si, the difference in the magnitudes of perpendicular and parallel MSRD becomes more evident, as the perpendicular vibrations can be represented as linear combinations of the two other basis vectors. Because the MSDs of the atoms have the same values for each cartesian component, it is expected that the relative displacements are distributed among the perpendicular and parallel MSRD as shown in Fig. \ref{fig:si-alpha-para-pen-01_02}.

Fig. \ref{fig:II-parallel-normalizedparal-pend-kp2} shows the normalized parallel and perpendicular MSRD values for the clathrate framework II. All results presented in Fig. \ref{fig:II-parallel-normalizedparal-pend-kp2} are for $\kappa'=2$ and $\kappa$ runs over all atoms within the same unit cell. The normalized parallel correlation functions have again similar values as in the previous works for \textit{d}-Si obtained using empirical potentials.\cite{0022-3719-13-13-005.Nielsen.Weber.displ.corr.1980} Both the parallel and perpendicular MSRD have similar characteristics, that is, the nearest atoms have the smallest MSRD values and vice versa. The atoms $\kappa=7-10$ have the smallest values and the atoms $\kappa=1,3$ the largest values in all cases in Fig. \ref{fig:II-parallel-normalizedparal-pend-kp2}. The atoms $\kappa=1,3$ are on the opposite side of the cavity with respect to atom $\kappa'=2$ (see Fig. \ref{fig:unit-cell-II}), which may partly explain the higher values of the MSRD in comparison to the other atoms. The perpendicular MSRD values do not form as distinct groups for the three symmetry equivalent atoms as in the case of the parallel MSRD.

\begin{figure}
\includegraphics[width=0.48\textwidth]{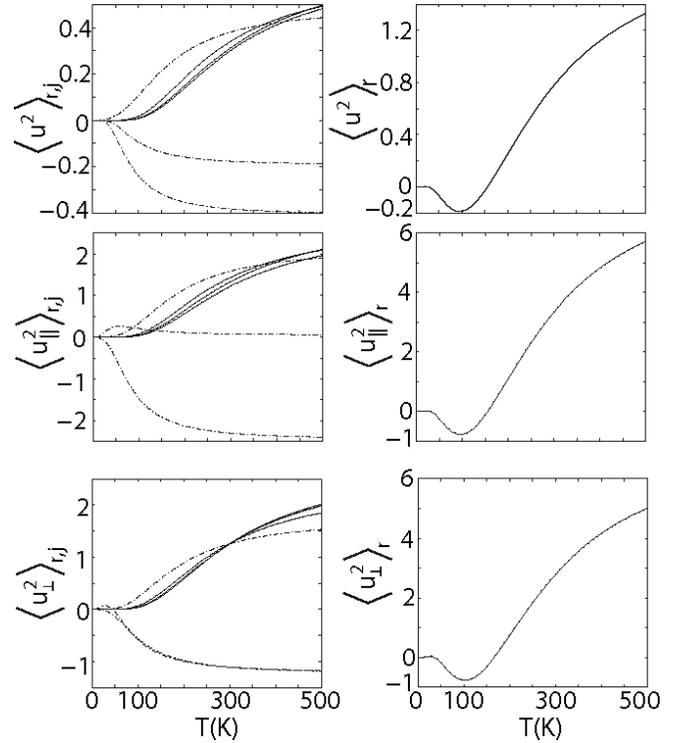}
\caption{MSD-VD- and MSRD-VD values for \textit{d}-Si calculated from Eq. \ref{eq:deltatotalMSD1}, Eq. \ref{eq:deltatotalparal}, and Eq. \ref{eq:deltatotalMSD1perdendic}. The MSRD-VD values are for the nearest neighbours. The acoustic modes are indicated with a dash-dotted line.} 
\label{fig:si-alpha-delta-MSD-MSRD-paper}
\end{figure}

Fig. \ref{fig:si-alpha-delta-MSD-MSRD-paper} shows the MSD-VD and MSRD-VD values for \textit{d}-Si. The values in Fig. \ref{fig:si-alpha-delta-MSD-MSRD-paper} for the different modes and for the total volume derivatives have similar characteristics as CTE calculated from Eq. \ref{eq:alphav-entrspy}. 
The reason for this similarity was discussed at the end of Sec. \ref{corr.funct.}. The MSD-VD values for the phonon modes corresponding to NTE are negative within the whole temperature range similar to the entropies shown in Fig. \ref{fig:deltaS}. Like in the case of the entropy, the MSD-VD values are negative for two acoustic phonon modes, while one acoustic mode and the optical modes have a positive contribution to CTE. The similarity of the MSRD-VD and CTE results could be due to the fact that in \textit{d}-Si only the relative motion of the two atoms within the primitive unit cell atoms determines the CTE. According to our results, the parallel and perpendicular MSRD-VD have nearly identical values.

\begin{figure}
\includegraphics[width=0.48\textwidth]{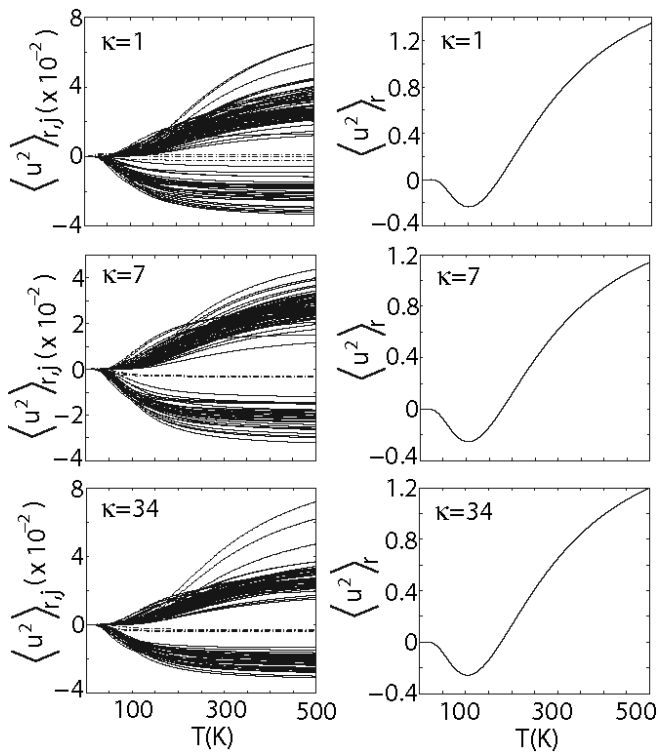}
\caption{MSD-VD for three different atoms in the structure II calculated from Eq. \ref{eq:deltatotalMSD1}. Acoustic modes are indicated with a dash-dotted line.} 
\label{fig:si-II-delta-MSD-paper}
\end{figure}

The MSD-VD values for the clathrate framework II shown in Fig. \ref{fig:si-II-delta-MSD-paper} correspond relatively well to the values for \textit{d}-Si.
Only three atoms representing the different symmetry independent positions in the crystal structure are shown. Comparison of Fig. \ref{fig:deltaS} and Fig. \ref{fig:si-II-delta-MSD-paper} implies that the MSD-VD and $\Delta S/\Delta V$ behave similarly as a function of temperature also for the structure II. The MSD-VD results also agree with the results obtained for CTE in the sense that the structure II has slightly lower MSD-VD values in both positive CTE and NTE temperature ranges when comparing it to \textit{d}-Si. The proper treatment of MSRD-VD values in larger structures requires symmetry considerations.

\section{Conclusions}
We have investigated the thermal properties of silicon clathrate frameworks by combining \textit{ab initio} DFT lattice dynamics with QHA. The computational results for \textit{d}-Si were in relatively good agreement with the experimental values for the phonon dispersion relations, heat capacities, and CTE. Studied clathrate frameworks have similar phonon spectra, only the structure VII differing from the others significantly. All studied clathrate frameworks were found to possess a phonon band gap and similar threshold frequencies confining the phonons that can have negative Gr\"{u}neisen parameter values. A similar threshold frequency was also found for \textit{d}-Si. Furthermore, the clathrate framework V has two phonon band gaps. The NTE temperature range of the studied clathrate frameworks is slightly wider than for \textit{d}-Si. The MSD and MSRD values calculated for \textit{d}-Si were in agreement with the previous computational results. The results obtained from the displacement correlation functions for the clathrate framework II were similar to \textit{d}-Si. Equal time displacement correlation functions were used to study the NTE phenomenon. The MSD-VD and MSRD-VD quantities defined here behave almost identically to the vibrational entropy volume derivatives in the studied structures. The parallel and perpendicular MSRD-VD results for \textit{d}-Si indicate that both quantities behave in similar fashion as a function of temperature. Further application of the volume derivatives of the displacement correlation functions may produce useful information on the mechanisms of the NTE phenomenon.

\begin{acknowledgments}
We thank Prof. Andrea Dal Corso (SISSA, Trieste) for providing modified QE routines to enable efficient distributed calculation of the phonon dispersion relations. We also thank Prof. Robert van Leeuwen and Dr. Gerrit Groenhof (University of Jyv\"{a}skyl\"{a}) for useful discussions on various aspects of the present work. The computing resources for this work were provided by the Finnish Grid Infrastructure (FGI). We gratefully acknowledge funding from the Foundation for Research of Natural Resources in Finland (grant 17591/13) and the Academy of Finland (grant 138560/2010).
\end{acknowledgments}

\bibliography{bibfile}

\begin{thebibliography}{53}%
\makeatletter
\providecommand \@ifxundefined [1]{%
 \@ifx{#1\undefined}
}%
\providecommand \@ifnum [1]{%
 \ifnum #1\expandafter \@firstoftwo
 \else \expandafter \@secondoftwo
 \fi
}%
\providecommand \@ifx [1]{%
 \ifx #1\expandafter \@firstoftwo
 \else \expandafter \@secondoftwo
 \fi
}%
\providecommand \natexlab [1]{#1}%
\providecommand \enquote  [1]{``#1''}%
\providecommand \bibnamefont  [1]{#1}%
\providecommand \bibfnamefont [1]{#1}%
\providecommand \citenamefont [1]{#1}%
\providecommand \href@noop [0]{\@secondoftwo}%
\providecommand \href [0]{\begingroup \@sanitize@url \@href}%
\providecommand \@href[1]{\@@startlink{#1}\@@href}%
\providecommand \@@href[1]{\endgroup#1\@@endlink}%
\providecommand \@sanitize@url [0]{\catcode `\\12\catcode `\$12\catcode
  `\&12\catcode `\#12\catcode `\^12\catcode `\_12\catcode `\%12\relax}%
\providecommand \@@startlink[1]{}%
\providecommand \@@endlink[0]{}%
\providecommand \url  [0]{\begingroup\@sanitize@url \@url }%
\providecommand \@url [1]{\endgroup\@href {#1}{\urlprefix }}%
\providecommand \urlprefix  [0]{URL }%
\providecommand \Eprint [0]{\href }%
\providecommand \doibase [0]{http://dx.doi.org/}%
\providecommand \selectlanguage [0]{\@gobble}%
\providecommand \bibinfo  [0]{\@secondoftwo}%
\providecommand \bibfield  [0]{\@secondoftwo}%
\providecommand \translation [1]{[#1]}%
\providecommand \BibitemOpen [0]{}%
\providecommand \bibitemStop [0]{}%
\providecommand \bibitemNoStop [0]{.\EOS\space}%
\providecommand \EOS [0]{\spacefactor3000\relax}%
\providecommand \BibitemShut  [1]{\csname bibitem#1\endcsname}%
\let\auto@bib@innerbib\@empty
\bibitem [{\citenamefont {Kasper}\ \emph {et~al.}(1965)\citenamefont {Kasper},
  \citenamefont {Hagenmuller}, \citenamefont {Pouchard},\ and\ \citenamefont
  {Cros}}]{Kasper24121965.science.clath.1965}%
  \BibitemOpen
  \bibfield  {author} {\bibinfo {author} {\bibfnamefont {J.~S.}\ \bibnamefont
  {Kasper}}, \bibinfo {author} {\bibfnamefont {P.}~\bibnamefont {Hagenmuller}},
  \bibinfo {author} {\bibfnamefont {M.}~\bibnamefont {Pouchard}}, \ and\
  \bibinfo {author} {\bibfnamefont {C.}~\bibnamefont {Cros}},\ }\href {\doibase
  10.1126/science.150.3704.1713} {\bibfield  {journal} {\bibinfo  {journal}
  {Science}\ }\textbf {\bibinfo {volume} {150}},\ \bibinfo {pages} {1713}
  (\bibinfo {year} {1965})}\BibitemShut {NoStop}%
\bibitem [{\citenamefont {Shevelkov}\ and\ \citenamefont
  {Kovnir}(2011)}]{Shevelkov-2011-zintl-clath}%
  \BibitemOpen
  \bibfield  {author} {\bibinfo {author} {\bibfnamefont {A.}~\bibnamefont
  {Shevelkov}}\ and\ \bibinfo {author} {\bibfnamefont {K.}~\bibnamefont
  {Kovnir}},\ }\href@noop {} {\emph {\bibinfo {title} {Zintl Clathrates}}}\
  (\bibinfo  {publisher} {Springer Berlin Heidelberg},\ \bibinfo {year}
  {2011})\ pp.\ \bibinfo {pages} {97--142}\BibitemShut {NoStop}%
\bibitem [{\citenamefont {Gryko}\ \emph {et~al.}(2000)\citenamefont {Gryko},
  \citenamefont {McMillan}, \citenamefont {Marzke}, \citenamefont
  {Ramachandran}, \citenamefont {Patton}, \citenamefont {Deb},\ and\
  \citenamefont {Sankey}}]{PhysRevB.62.R7707.empty.clathrates.Gryko}%
  \BibitemOpen
  \bibfield  {author} {\bibinfo {author} {\bibfnamefont {J.}~\bibnamefont
  {Gryko}}, \bibinfo {author} {\bibfnamefont {P.~F.}\ \bibnamefont {McMillan}},
  \bibinfo {author} {\bibfnamefont {R.~F.}\ \bibnamefont {Marzke}}, \bibinfo
  {author} {\bibfnamefont {G.~K.}\ \bibnamefont {Ramachandran}}, \bibinfo
  {author} {\bibfnamefont {D.}~\bibnamefont {Patton}}, \bibinfo {author}
  {\bibfnamefont {S.~K.}\ \bibnamefont {Deb}}, \ and\ \bibinfo {author}
  {\bibfnamefont {O.~F.}\ \bibnamefont {Sankey}},\ }\href {\doibase
  10.1103/PhysRevB.62.R7707} {\bibfield  {journal} {\bibinfo  {journal} {Phys.
  Rev. B}\ }\textbf {\bibinfo {volume} {62}},\ \bibinfo {pages} {R7707}
  (\bibinfo {year} {2000})}\BibitemShut {NoStop}%
\bibitem [{\citenamefont {Ammar}\ \emph {et~al.}(2004)\citenamefont {Ammar},
  \citenamefont {Cros}, \citenamefont {Pouchard}, \citenamefont {Jaussaud},
  \citenamefont {Bassat}, \citenamefont {Villeneuve}, \citenamefont {Duttine},
  \citenamefont {Ménétrier},\ and\ \citenamefont
  {Reny}}]{Ammar2004393-clath}%
  \BibitemOpen
  \bibfield  {author} {\bibinfo {author} {\bibfnamefont {A.}~\bibnamefont
  {Ammar}}, \bibinfo {author} {\bibfnamefont {C.}~\bibnamefont {Cros}},
  \bibinfo {author} {\bibfnamefont {M.}~\bibnamefont {Pouchard}}, \bibinfo
  {author} {\bibfnamefont {N.}~\bibnamefont {Jaussaud}}, \bibinfo {author}
  {\bibfnamefont {J.-M.}\ \bibnamefont {Bassat}}, \bibinfo {author}
  {\bibfnamefont {G.}~\bibnamefont {Villeneuve}}, \bibinfo {author}
  {\bibfnamefont {M.}~\bibnamefont {Duttine}}, \bibinfo {author} {\bibfnamefont
  {M.}~\bibnamefont {Ménétrier}}, \ and\ \bibinfo {author} {\bibfnamefont
  {E.}~\bibnamefont {Reny}},\ }\href {\doibase
  http://dx.doi.org/10.1016/j.solidstatesciences.2004.02.006} {\bibfield
  {journal} {\bibinfo  {journal} {Solid State Sci.}\ }\textbf {\bibinfo
  {volume} {6}},\ \bibinfo {pages} {393 } (\bibinfo {year} {2004})}\BibitemShut
  {NoStop}%
\bibitem [{\citenamefont {Adams}\ \emph {et~al.}(1994)\citenamefont {Adams},
  \citenamefont {O'Keeffe}, \citenamefont {Demkov}, \citenamefont {Sankey},\
  and\ \citenamefont {Huang}}]{Adams.et.al.1994-PhysRevB.49.8048}%
  \BibitemOpen
  \bibfield  {author} {\bibinfo {author} {\bibfnamefont {G.~B.}\ \bibnamefont
  {Adams}}, \bibinfo {author} {\bibfnamefont {M.}~\bibnamefont {O'Keeffe}},
  \bibinfo {author} {\bibfnamefont {A.~A.}\ \bibnamefont {Demkov}}, \bibinfo
  {author} {\bibfnamefont {O.~F.}\ \bibnamefont {Sankey}}, \ and\ \bibinfo
  {author} {\bibfnamefont {Y.-M.}\ \bibnamefont {Huang}},\ }\href {\doibase
  10.1103/PhysRevB.49.8048} {\bibfield  {journal} {\bibinfo  {journal} {Phys.
  Rev. B}\ }\textbf {\bibinfo {volume} {49}},\ \bibinfo {pages} {8048}
  (\bibinfo {year} {1994})}\BibitemShut {NoStop}%
\bibitem [{\citenamefont {Saito}\ and\ \citenamefont
  {Oshiyama}(1995)}]{Saito.et.al.electr.calc.clathI-PhysRevB.51.2628-1995}%
  \BibitemOpen
  \bibfield  {author} {\bibinfo {author} {\bibfnamefont {S.}~\bibnamefont
  {Saito}}\ and\ \bibinfo {author} {\bibfnamefont {A.}~\bibnamefont
  {Oshiyama}},\ }\href {\doibase 10.1103/PhysRevB.51.2628} {\bibfield
  {journal} {\bibinfo  {journal} {Phys. Rev. B}\ }\textbf {\bibinfo {volume}
  {51}},\ \bibinfo {pages} {2628} (\bibinfo {year} {1995})}\BibitemShut
  {NoStop}%
\bibitem [{\citenamefont {Kawaji}\ \emph {et~al.}(1995)\citenamefont {Kawaji},
  \citenamefont {Horie}, \citenamefont {Yamanaka},\ and\ \citenamefont
  {Ishikawa}}]{PhysRevLett.74.1427.Si.clath.superconductivity}%
  \BibitemOpen
  \bibfield  {author} {\bibinfo {author} {\bibfnamefont {H.}~\bibnamefont
  {Kawaji}}, \bibinfo {author} {\bibfnamefont {H.-o.}\ \bibnamefont {Horie}},
  \bibinfo {author} {\bibfnamefont {S.}~\bibnamefont {Yamanaka}}, \ and\
  \bibinfo {author} {\bibfnamefont {M.}~\bibnamefont {Ishikawa}},\ }\href
  {\doibase 10.1103/PhysRevLett.74.1427} {\bibfield  {journal} {\bibinfo
  {journal} {Phys. Rev. Lett.}\ }\textbf {\bibinfo {volume} {74}},\ \bibinfo
  {pages} {1427} (\bibinfo {year} {1995})}\BibitemShut {NoStop}%
\bibitem [{\citenamefont {Nolas}\ \emph {et~al.}(1998)\citenamefont {Nolas},
  \citenamefont {Cohn}, \citenamefont {Slack},\ and\ \citenamefont
  {Schujman}}]{4899409Nolas-1998-Ge-clath-thermoelect.}%
  \BibitemOpen
  \bibfield  {author} {\bibinfo {author} {\bibfnamefont {G.}~\bibnamefont
  {Nolas}}, \bibinfo {author} {\bibfnamefont {J.}~\bibnamefont {Cohn}},
  \bibinfo {author} {\bibfnamefont {G.}~\bibnamefont {Slack}}, \ and\ \bibinfo
  {author} {\bibfnamefont {S.}~\bibnamefont {Schujman}},\ }\href {\doibase
  10.1063/1.121747} {\bibfield  {journal} {\bibinfo  {journal} {Appl. Phys.
  Lett.}\ }\textbf {\bibinfo {volume} {73}},\ \bibinfo {pages} {178} (\bibinfo
  {year} {1998})}\BibitemShut {NoStop}%
\bibitem [{\citenamefont {Sootsman}\ \emph {et~al.}(2009)\citenamefont
  {Sootsman}, \citenamefont {Chung},\ and\ \citenamefont
  {Kanatzidis}}]{Sootsman-thermoelect.-review.2009-ANIE200900598}%
  \BibitemOpen
  \bibfield  {author} {\bibinfo {author} {\bibfnamefont {J.}~\bibnamefont
  {Sootsman}}, \bibinfo {author} {\bibfnamefont {D.}~\bibnamefont {Chung}}, \
  and\ \bibinfo {author} {\bibfnamefont {M.}~\bibnamefont {Kanatzidis}},\
  }\href {\doibase 10.1002/anie.200900598} {\bibfield  {journal} {\bibinfo
  {journal} {Angew. Chem. Int. Ed.}\ }\textbf {\bibinfo {volume} {48}},\
  \bibinfo {pages} {8616} (\bibinfo {year} {2009})}\BibitemShut {NoStop}%
\bibitem [{\citenamefont {Christensen}\ \emph {et~al.}(2010)\citenamefont
  {Christensen}, \citenamefont {Johnsen},\ and\ \citenamefont
  {Iversen}}]{Christensen.et.al.-2010-B916400F}%
  \BibitemOpen
  \bibfield  {author} {\bibinfo {author} {\bibfnamefont {M.}~\bibnamefont
  {Christensen}}, \bibinfo {author} {\bibfnamefont {S.}~\bibnamefont
  {Johnsen}}, \ and\ \bibinfo {author} {\bibfnamefont {B.~B.}\ \bibnamefont
  {Iversen}},\ }\href {\doibase 10.1039/B916400F} {\bibfield  {journal}
  {\bibinfo  {journal} {Dalton Trans.}\ }\textbf {\bibinfo {volume} {39}},\
  \bibinfo {pages} {978} (\bibinfo {year} {2010})}\BibitemShut {NoStop}%
\bibitem [{\citenamefont {Kuznetsov}\ \emph {et~al.}(2000)\citenamefont
  {Kuznetsov}, \citenamefont {Kuznetsova}, \citenamefont {Kaliazin},\ and\
  \citenamefont {Rowe}}]{Kuznetsov-et.al.2000-jap/87/11/10.1063/1.373469}%
  \BibitemOpen
  \bibfield  {author} {\bibinfo {author} {\bibfnamefont {V.~L.}\ \bibnamefont
  {Kuznetsov}}, \bibinfo {author} {\bibfnamefont {L.~A.}\ \bibnamefont
  {Kuznetsova}}, \bibinfo {author} {\bibfnamefont {A.~E.}\ \bibnamefont
  {Kaliazin}}, \ and\ \bibinfo {author} {\bibfnamefont {D.~M.}\ \bibnamefont
  {Rowe}},\ }\href {\doibase http://dx.doi.org/10.1063/1.373469} {\bibfield
  {journal} {\bibinfo  {journal} {J. Appl. Phys.}\ }\textbf {\bibinfo {volume}
  {87}},\ \bibinfo {pages} {7871} (\bibinfo {year} {2000})}\BibitemShut
  {NoStop}%
\bibitem [{\citenamefont {Toberer}\ \emph {et~al.}(2008)\citenamefont
  {Toberer}, \citenamefont {Christensen}, \citenamefont {Iversen},\ and\
  \citenamefont {Snyder}}]{Toberer.et.al-2008-PhysRevB.77.075203}%
  \BibitemOpen
  \bibfield  {author} {\bibinfo {author} {\bibfnamefont {E.~S.}\ \bibnamefont
  {Toberer}}, \bibinfo {author} {\bibfnamefont {M.}~\bibnamefont
  {Christensen}}, \bibinfo {author} {\bibfnamefont {B.~B.}\ \bibnamefont
  {Iversen}}, \ and\ \bibinfo {author} {\bibfnamefont {G.~J.}\ \bibnamefont
  {Snyder}},\ }\href {\doibase 10.1103/PhysRevB.77.075203} {\bibfield
  {journal} {\bibinfo  {journal} {Phys. Rev. B}\ }\textbf {\bibinfo {volume}
  {77}},\ \bibinfo {pages} {075203} (\bibinfo {year} {2008})}\BibitemShut
  {NoStop}%
\bibitem [{\citenamefont {Tang}\ \emph {et~al.}(2006)\citenamefont {Tang},
  \citenamefont {Dong}, \citenamefont {Hutchins}, \citenamefont {Shebanova},
  \citenamefont {Gryko}, \citenamefont {Barnes}, \citenamefont {Cockcroft},
  \citenamefont {Vickers},\ and\ \citenamefont
  {McMillan}}]{PhysRevB.74.014109.Tang.et.al.thermal.prop.clathII}%
  \BibitemOpen
  \bibfield  {author} {\bibinfo {author} {\bibfnamefont {X.}~\bibnamefont
  {Tang}}, \bibinfo {author} {\bibfnamefont {J.}~\bibnamefont {Dong}}, \bibinfo
  {author} {\bibfnamefont {P.}~\bibnamefont {Hutchins}}, \bibinfo {author}
  {\bibfnamefont {O.}~\bibnamefont {Shebanova}}, \bibinfo {author}
  {\bibfnamefont {J.}~\bibnamefont {Gryko}}, \bibinfo {author} {\bibfnamefont
  {P.}~\bibnamefont {Barnes}}, \bibinfo {author} {\bibfnamefont {J.~K.}\
  \bibnamefont {Cockcroft}}, \bibinfo {author} {\bibfnamefont {M.}~\bibnamefont
  {Vickers}}, \ and\ \bibinfo {author} {\bibfnamefont {P.~F.}\ \bibnamefont
  {McMillan}},\ }\href {\doibase 10.1103/PhysRevB.74.014109} {\bibfield
  {journal} {\bibinfo  {journal} {Phys. Rev. B}\ }\textbf {\bibinfo {volume}
  {74}},\ \bibinfo {pages} {014109} (\bibinfo {year} {2006})}\BibitemShut
  {NoStop}%
\bibitem [{\citenamefont {Evans}\ \emph {et~al.}(1997)\citenamefont {Evans},
  \citenamefont {Mary},\ and\ \citenamefont {Sleight}}]{Evans1997311-NTE-1998}%
  \BibitemOpen
  \bibfield  {author} {\bibinfo {author} {\bibfnamefont {J.}~\bibnamefont
  {Evans}}, \bibinfo {author} {\bibfnamefont {T.}~\bibnamefont {Mary}}, \ and\
  \bibinfo {author} {\bibfnamefont {A.}~\bibnamefont {Sleight}},\ }\href
  {\doibase http://dx.doi.org/10.1016/S0921-4526(97)00571-1} {\bibfield
  {journal} {\bibinfo  {journal} {Physica B}\ }\textbf {\bibinfo {volume}
  {241-–243}},\ \bibinfo {pages} {311 } (\bibinfo {year} {1997})}\BibitemShut
  {NoStop}%
\bibitem [{\citenamefont {Evans}\ \emph {et~al.}(1996)\citenamefont {Evans},
  \citenamefont {Mary}, \citenamefont {Vogt}, \citenamefont {Subramanian},\
  and\ \citenamefont {Sleight}}]{chem.mat.9602959.Evans.Mary.1996.NTE}%
  \BibitemOpen
  \bibfield  {author} {\bibinfo {author} {\bibfnamefont {J.~S.~O.}\
  \bibnamefont {Evans}}, \bibinfo {author} {\bibfnamefont {T.~A.}\ \bibnamefont
  {Mary}}, \bibinfo {author} {\bibfnamefont {T.}~\bibnamefont {Vogt}}, \bibinfo
  {author} {\bibfnamefont {M.~A.}\ \bibnamefont {Subramanian}}, \ and\ \bibinfo
  {author} {\bibfnamefont {A.~W.}\ \bibnamefont {Sleight}},\ }\href {\doibase
  10.1021/cm9602959} {\bibfield  {journal} {\bibinfo  {journal} {Chem. Mater.}\
  }\textbf {\bibinfo {volume} {8}},\ \bibinfo {pages} {2809} (\bibinfo {year}
  {1996})}\BibitemShut {NoStop}%
\bibitem [{\citenamefont {Mary}\ \emph {et~al.}(1996)\citenamefont {Mary},
  \citenamefont {Evans}, \citenamefont {Vogt},\ and\ \citenamefont
  {Sleight}}]{Mary05041996.science.ZrW2O8.NTE}%
  \BibitemOpen
  \bibfield  {author} {\bibinfo {author} {\bibfnamefont {T.~A.}\ \bibnamefont
  {Mary}}, \bibinfo {author} {\bibfnamefont {J.~S.~O.}\ \bibnamefont {Evans}},
  \bibinfo {author} {\bibfnamefont {T.}~\bibnamefont {Vogt}}, \ and\ \bibinfo
  {author} {\bibfnamefont {A.~W.}\ \bibnamefont {Sleight}},\ }\href {\doibase
  10.1126/science.272.5258.90} {\bibfield  {journal} {\bibinfo  {journal}
  {Science}\ }\textbf {\bibinfo {volume} {272}},\ \bibinfo {pages} {90}
  (\bibinfo {year} {1996})}\BibitemShut {NoStop}%
\bibitem [{\citenamefont {Barrera}\ \emph {et~al.}(2005)\citenamefont
  {Barrera}, \citenamefont {Bruno}, \citenamefont {Barron},\ and\ \citenamefont
  {Allan}}]{0953-8984-17-4-R03.jour.phys.cond.mat..Barrera.Bruno.Barron.NTE-rew.}%
  \BibitemOpen
  \bibfield  {author} {\bibinfo {author} {\bibfnamefont {G.~D.}\ \bibnamefont
  {Barrera}}, \bibinfo {author} {\bibfnamefont {J.~A.~O.}\ \bibnamefont
  {Bruno}}, \bibinfo {author} {\bibfnamefont {T.~H.~K.}\ \bibnamefont
  {Barron}}, \ and\ \bibinfo {author} {\bibfnamefont {N.~L.}\ \bibnamefont
  {Allan}},\ }\href {http://stacks.iop.org/0953-8984/17/i=4/a=R03} {\bibfield
  {journal} {\bibinfo  {journal} {J. Phys.}\ }\textbf {\bibinfo {volume}
  {17}},\ \bibinfo {pages} {R217} (\bibinfo {year} {2005})}\BibitemShut
  {NoStop}%
\bibitem [{\citenamefont {Reeber}\ and\ \citenamefont
  {Wang}(1996)}]{Reeber1996259.Sialpha.thermalexp.experimental}%
  \BibitemOpen
  \bibfield  {author} {\bibinfo {author} {\bibfnamefont {R.~R.}\ \bibnamefont
  {Reeber}}\ and\ \bibinfo {author} {\bibfnamefont {K.}~\bibnamefont {Wang}},\
  }\href {\doibase http://dx.doi.org/10.1016/S0254-0584(96)01808-1} {\bibfield
  {journal} {\bibinfo  {journal} {Mater. Chem. Phys.}\ }\textbf {\bibinfo
  {volume} {46}},\ \bibinfo {pages} {259 } (\bibinfo {year}
  {1996})}\BibitemShut {NoStop}%
\bibitem [{\citenamefont {Sayers}\ \emph {et~al.}(1971)\citenamefont {Sayers},
  \citenamefont {Stern},\ and\ \citenamefont
  {Lytle}}]{PhysRevLett.27.1204.Sayers.et.al.1971.EXAFS}%
  \BibitemOpen
  \bibfield  {author} {\bibinfo {author} {\bibfnamefont {D.~E.}\ \bibnamefont
  {Sayers}}, \bibinfo {author} {\bibfnamefont {E.~A.}\ \bibnamefont {Stern}}, \
  and\ \bibinfo {author} {\bibfnamefont {F.~W.}\ \bibnamefont {Lytle}},\ }\href
  {\doibase 10.1103/PhysRevLett.27.1204} {\bibfield  {journal} {\bibinfo
  {journal} {Phys. Rev. Lett.}\ }\textbf {\bibinfo {volume} {27}},\ \bibinfo
  {pages} {1204} (\bibinfo {year} {1971})}\BibitemShut {NoStop}%
\bibitem [{\citenamefont {Lee}\ and\ \citenamefont
  {Pendry}(1975)}]{PhysRevB.11.2795.Lee.et.al.EXAFS.1975}%
  \BibitemOpen
  \bibfield  {author} {\bibinfo {author} {\bibfnamefont {P.~A.}\ \bibnamefont
  {Lee}}\ and\ \bibinfo {author} {\bibfnamefont {J.~B.}\ \bibnamefont
  {Pendry}},\ }\href {\doibase 10.1103/PhysRevB.11.2795} {\bibfield  {journal}
  {\bibinfo  {journal} {Phys. Rev. B}\ }\textbf {\bibinfo {volume} {11}},\
  \bibinfo {pages} {2795} (\bibinfo {year} {1975})}\BibitemShut {NoStop}%
\bibitem [{\citenamefont {Hayes}\ \emph {et~al.}(1976)\citenamefont {Hayes},
  \citenamefont {Sen},\ and\ \citenamefont
  {Hunter}}]{0022-3719-9-24-006.Hayes.et.al.EXAFS1976}%
  \BibitemOpen
  \bibfield  {author} {\bibinfo {author} {\bibfnamefont {T.~M.}\ \bibnamefont
  {Hayes}}, \bibinfo {author} {\bibfnamefont {P.~N.}\ \bibnamefont {Sen}}, \
  and\ \bibinfo {author} {\bibfnamefont {S.~H.}\ \bibnamefont {Hunter}},\
  }\href {http://stacks.iop.org/0022-3719/9/i=24/a=006} {\bibfield  {journal}
  {\bibinfo  {journal} {J. Phys. C}\ }\textbf {\bibinfo {volume} {9}},\
  \bibinfo {pages} {4357} (\bibinfo {year} {1976})}\BibitemShut {NoStop}%
\bibitem [{\citenamefont {Tr\"oger}\ \emph {et~al.}(1994)\citenamefont
  {Tr\"oger}, \citenamefont {Yokoyama}, \citenamefont {Arvanitis},
  \citenamefont {Lederer}, \citenamefont {Tischer},\ and\ \citenamefont
  {Baberschke}}]{PhysRevB.49.888.Troger.et.al.EXAFS.1994}%
  \BibitemOpen
  \bibfield  {author} {\bibinfo {author} {\bibfnamefont {L.}~\bibnamefont
  {Tr\"oger}}, \bibinfo {author} {\bibfnamefont {T.}~\bibnamefont {Yokoyama}},
  \bibinfo {author} {\bibfnamefont {D.}~\bibnamefont {Arvanitis}}, \bibinfo
  {author} {\bibfnamefont {T.}~\bibnamefont {Lederer}}, \bibinfo {author}
  {\bibfnamefont {M.}~\bibnamefont {Tischer}}, \ and\ \bibinfo {author}
  {\bibfnamefont {K.}~\bibnamefont {Baberschke}},\ }\href {\doibase
  10.1103/PhysRevB.49.888} {\bibfield  {journal} {\bibinfo  {journal} {Phys.
  Rev. B}\ }\textbf {\bibinfo {volume} {49}},\ \bibinfo {pages} {888} (\bibinfo
  {year} {1994})}\BibitemShut {NoStop}%
\bibitem [{\citenamefont {Sanson}\ \emph {et~al.}(2006)\citenamefont {Sanson},
  \citenamefont {Rocca}, \citenamefont {Dalba}, \citenamefont {Fornasini},
  \citenamefont {Grisenti}, \citenamefont {Dapiaggi},\ and\ \citenamefont
  {Artioli}}]{PhysRevB.73.214305.Sanson.et.al.EXAFS.NTE.2006}%
  \BibitemOpen
  \bibfield  {author} {\bibinfo {author} {\bibfnamefont {A.}~\bibnamefont
  {Sanson}}, \bibinfo {author} {\bibfnamefont {F.}~\bibnamefont {Rocca}},
  \bibinfo {author} {\bibfnamefont {G.}~\bibnamefont {Dalba}}, \bibinfo
  {author} {\bibfnamefont {P.}~\bibnamefont {Fornasini}}, \bibinfo {author}
  {\bibfnamefont {R.}~\bibnamefont {Grisenti}}, \bibinfo {author}
  {\bibfnamefont {M.}~\bibnamefont {Dapiaggi}}, \ and\ \bibinfo {author}
  {\bibfnamefont {G.}~\bibnamefont {Artioli}},\ }\href {\doibase
  10.1103/PhysRevB.73.214305} {\bibfield  {journal} {\bibinfo  {journal} {Phys.
  Rev. B}\ }\textbf {\bibinfo {volume} {73}},\ \bibinfo {pages} {214305}
  (\bibinfo {year} {2006})}\BibitemShut {NoStop}%
\bibitem [{\citenamefont {el~All}\ \emph {et~al.}(2012)\citenamefont {el~All},
  \citenamefont {Dalba}, \citenamefont {Diop}, \citenamefont {Fornasini},
  \citenamefont {Grisenti}, \citenamefont {Mathon}, \citenamefont {Rocca},
  \citenamefont {Sendja},\ and\ \citenamefont
  {Vaccari}}]{0953-8984-24-11-115403.Abd.et.al.EXAFS.NTE.2012}%
  \BibitemOpen
  \bibfield  {author} {\bibinfo {author} {\bibfnamefont {N.~A.}\ \bibnamefont
  {el~All}}, \bibinfo {author} {\bibfnamefont {G.}~\bibnamefont {Dalba}},
  \bibinfo {author} {\bibfnamefont {D.}~\bibnamefont {Diop}}, \bibinfo {author}
  {\bibfnamefont {P.}~\bibnamefont {Fornasini}}, \bibinfo {author}
  {\bibfnamefont {R.}~\bibnamefont {Grisenti}}, \bibinfo {author}
  {\bibfnamefont {O.}~\bibnamefont {Mathon}}, \bibinfo {author} {\bibfnamefont
  {F.}~\bibnamefont {Rocca}}, \bibinfo {author} {\bibfnamefont {B.~T.}\
  \bibnamefont {Sendja}}, \ and\ \bibinfo {author} {\bibfnamefont
  {M.}~\bibnamefont {Vaccari}},\ }\href
  {http://stacks.iop.org/0953-8984/24/i=11/a=115403} {\bibfield  {journal}
  {\bibinfo  {journal} {J. Phys.}\ }\textbf {\bibinfo {volume} {24}},\ \bibinfo
  {pages} {115403} (\bibinfo {year} {2012})}\BibitemShut {NoStop}%
\bibitem [{\citenamefont {Huang}\ and\ \citenamefont
  {Born}(1954)}]{born-huang1954dynamical}%
  \BibitemOpen
  \bibfield  {author} {\bibinfo {author} {\bibfnamefont {K.}~\bibnamefont
  {Huang}}\ and\ \bibinfo {author} {\bibfnamefont {M.}~\bibnamefont {Born}},\
  }\href@noop {} {\emph {\bibinfo {title} {Dynamical Theory of Crystal
  Lattices}}}\ (\bibinfo  {publisher} {Clarendon Press Oxford},\ \bibinfo
  {year} {1954})\ p.\ \bibinfo {pages} {298}\BibitemShut {NoStop}%
\bibitem [{\citenamefont {Baroni}\ \emph {et~al.}(2001)\citenamefont {Baroni},
  \citenamefont {de~Gironcoli}, \citenamefont {Dal~Corso},\ and\ \citenamefont
  {Giannozzi}}]{RevModPhys.73.515-Baroni-2001-DFTP}%
  \BibitemOpen
  \bibfield  {author} {\bibinfo {author} {\bibfnamefont {S.}~\bibnamefont
  {Baroni}}, \bibinfo {author} {\bibfnamefont {S.}~\bibnamefont
  {de~Gironcoli}}, \bibinfo {author} {\bibfnamefont {A.}~\bibnamefont
  {Dal~Corso}}, \ and\ \bibinfo {author} {\bibfnamefont {P.}~\bibnamefont
  {Giannozzi}},\ }\href {\doibase 10.1103/RevModPhys.73.515} {\bibfield
  {journal} {\bibinfo  {journal} {Rev. Mod. Phys.}\ }\textbf {\bibinfo {volume}
  {73}},\ \bibinfo {pages} {515} (\bibinfo {year} {2001})}\BibitemShut
  {NoStop}%
\bibitem [{\citenamefont {Maradudin}\ \emph {et~al.}(1971)\citenamefont
  {Maradudin}, \citenamefont {Montroll}, \citenamefont {Weiss},\ and\
  \citenamefont {Ipatova}}]{maradudin1971-harm-appr}%
  \BibitemOpen
  \bibfield  {author} {\bibinfo {author} {\bibfnamefont {A.}~\bibnamefont
  {Maradudin}}, \bibinfo {author} {\bibfnamefont {E.~W.}\ \bibnamefont
  {Montroll}}, \bibinfo {author} {\bibfnamefont {G.~H.}\ \bibnamefont {Weiss}},
  \ and\ \bibinfo {author} {\bibfnamefont {I.~P.}\ \bibnamefont {Ipatova}},\
  }\href@noop {} {\emph {\bibinfo {title} {Theory of The Lattice Dynamics in
  The Harmonic Approximation}}},\ Vol.\ \bibinfo {volume} {Supplement 3}\
  (\bibinfo  {publisher} {Academic Press},\ \bibinfo {year} {1971})\
  p.~\bibinfo {pages} {62}\BibitemShut {NoStop}%
\bibitem [{\citenamefont {Ashcroft}\ and\ \citenamefont
  {Mermin}(1976)}]{ashcroft-solid-state-phys}%
  \BibitemOpen
  \bibfield  {author} {\bibinfo {author} {\bibfnamefont {N.}~\bibnamefont
  {Ashcroft}}\ and\ \bibinfo {author} {\bibfnamefont {D.}~\bibnamefont
  {Mermin}},\ }\href@noop {} {\emph {\bibinfo {title} {Solid State Physics}}}\
  (\bibinfo  {publisher} {Harcourt},\ \bibinfo {year} {1976})\BibitemShut
  {NoStop}%
\bibitem [{\citenamefont {Taylor}(1998)}]{taylor1998thermal}%
  \BibitemOpen
  \bibfield  {author} {\bibinfo {author} {\bibfnamefont {R.~E.}\ \bibnamefont
  {Taylor}},\ }\href@noop {} {\emph {\bibinfo {title} {Thermal Expansion of
  Solids}}},\ Vol.~\bibinfo {volume} {4}\ (\bibinfo  {publisher} {ASM
  international},\ \bibinfo {year} {1998})\BibitemShut {NoStop}%
\bibitem [{\citenamefont {Beni}\ and\ \citenamefont
  {Platzman}(1976)}]{PhysRevB.14.1514.Beni.EXAFS1976}%
  \BibitemOpen
  \bibfield  {author} {\bibinfo {author} {\bibfnamefont {G.}~\bibnamefont
  {Beni}}\ and\ \bibinfo {author} {\bibfnamefont {P.~M.}\ \bibnamefont
  {Platzman}},\ }\href {\doibase 10.1103/PhysRevB.14.1514} {\bibfield
  {journal} {\bibinfo  {journal} {Phys. Rev. B}\ }\textbf {\bibinfo {volume}
  {14}},\ \bibinfo {pages} {1514} (\bibinfo {year} {1976})}\BibitemShut
  {NoStop}%
\bibitem [{\citenamefont
  {Fornasini}(2001)}]{0953-8984-13-34-324-Fornasini2001}%
  \BibitemOpen
  \bibfield  {author} {\bibinfo {author} {\bibfnamefont {P.}~\bibnamefont
  {Fornasini}},\ }\href {http://stacks.iop.org/0953-8984/13/i=34/a=324}
  {\bibfield  {journal} {\bibinfo  {journal} {J. Phys.}\ }\textbf {\bibinfo
  {volume} {13}},\ \bibinfo {pages} {7859} (\bibinfo {year}
  {2001})}\BibitemShut {NoStop}%
\bibitem [{\citenamefont {Nielsen}\ and\ \citenamefont
  {Weber}(1980)}]{0022-3719-13-13-005.Nielsen.Weber.displ.corr.1980}%
  \BibitemOpen
  \bibfield  {author} {\bibinfo {author} {\bibfnamefont {O.~H.}\ \bibnamefont
  {Nielsen}}\ and\ \bibinfo {author} {\bibfnamefont {W.}~\bibnamefont
  {Weber}},\ }\href {http://stacks.iop.org/0022-3719/13/i=13/a=005} {\bibfield
  {journal} {\bibinfo  {journal} {J. Phys. C}\ }\textbf {\bibinfo {volume}
  {13}},\ \bibinfo {pages} {2449} (\bibinfo {year} {1980})}\BibitemShut
  {NoStop}%
\bibitem [{\citenamefont
  {Weber}(1987)}]{0305-4608-17-1-009.Weber.displ.corr.1987}%
  \BibitemOpen
  \bibfield  {author} {\bibinfo {author} {\bibfnamefont {W.}~\bibnamefont
  {Weber}},\ }\href {http://stacks.iop.org/0305-4608/17/i=1/a=009} {\bibfield
  {journal} {\bibinfo  {journal} {J. Phys. F: Metal Phys.}\ }\textbf {\bibinfo
  {volume} {17}},\ \bibinfo {pages} {27} (\bibinfo {year} {1987})}\BibitemShut
  {NoStop}%
\bibitem [{\citenamefont {Karttunen}\ \emph {et~al.}(2010)\citenamefont
  {Karttunen}, \citenamefont {Fässler}, \citenamefont {Linnolahti},\ and\
  \citenamefont {Pakkanen}}]{karttunen2010structuralprinc}%
  \BibitemOpen
  \bibfield  {author} {\bibinfo {author} {\bibfnamefont {A.~J.}\ \bibnamefont
  {Karttunen}}, \bibinfo {author} {\bibfnamefont {T.~F.}\ \bibnamefont
  {Fässler}}, \bibinfo {author} {\bibfnamefont {M.}~\bibnamefont
  {Linnolahti}}, \ and\ \bibinfo {author} {\bibfnamefont {T.~A.}\ \bibnamefont
  {Pakkanen}},\ }\href@noop {} {\bibfield  {journal} {\bibinfo  {journal}
  {Inorg. Chem.}\ }\textbf {\bibinfo {volume} {50}},\ \bibinfo {pages} {1733}
  (\bibinfo {year} {2010})}\BibitemShut {NoStop}%
\bibitem [{\citenamefont {Rogl}(2006)}]{Rogl2006}%
  \BibitemOpen
  \bibfield  {author} {\bibinfo {author} {\bibfnamefont {P.}~\bibnamefont
  {Rogl}},\ }\href@noop {} {\emph {\bibinfo {title} {Thermoelectrics Handbook:
  Macro to Nano}}}\ (\bibinfo  {publisher} {CRC Press},\ \bibinfo {year}
  {2006})\ pp.\ \bibinfo {pages} {32--1}\BibitemShut {NoStop}%
\bibitem [{\citenamefont {Kovnir}\ and\ \citenamefont
  {Shevelkov}(2004)}]{0036-021X-73-9-R06.Kovnir.Shevelkov.clath.review.2004.}%
  \BibitemOpen
  \bibfield  {author} {\bibinfo {author} {\bibfnamefont {K.~A.}\ \bibnamefont
  {Kovnir}}\ and\ \bibinfo {author} {\bibfnamefont {A.~V.}\ \bibnamefont
  {Shevelkov}},\ }\href {http://stacks.iop.org/0036-021X/73/i=9/a=R06}
  {\bibfield  {journal} {\bibinfo  {journal} {Russ. Chem. Rev.}\ }\textbf
  {\bibinfo {volume} {73}},\ \bibinfo {pages} {923} (\bibinfo {year}
  {2004})}\BibitemShut {NoStop}%
\bibitem [{\citenamefont {Giannozzi}\ \emph {et~al.}(2009)\citenamefont
  {Giannozzi}, \citenamefont {Baroni}, \citenamefont {Bonini}, \citenamefont
  {Calandra}, \citenamefont {Car}, \citenamefont {Cavazzoni}, \citenamefont
  {Ceresoli}, \citenamefont {Chiarotti}, \citenamefont {Cococcioni},
  \citenamefont {Dabo}, \citenamefont {{Dal Corso}}, \citenamefont
  {de~Gironcoli}, \citenamefont {Fabris}, \citenamefont {Fratesi},
  \citenamefont {Gebauer}, \citenamefont {Gerstmann}, \citenamefont
  {Gougoussis}, \citenamefont {Kokalj}, \citenamefont {Lazzeri}, \citenamefont
  {Martin-Samos}, \citenamefont {Marzari}, \citenamefont {Mauri}, \citenamefont
  {Mazzarello}, \citenamefont {Paolini}, \citenamefont {Pasquarello},
  \citenamefont {Paulatto}, \citenamefont {Sbraccia}, \citenamefont {Scandolo},
  \citenamefont {Sclauzero}, \citenamefont {Seitsonen}, \citenamefont
  {Smogunov}, \citenamefont {Umari},\ and\ \citenamefont
  {Wentzcovitch}}]{QE-2009}%
  \BibitemOpen
  \bibfield  {author} {\bibinfo {author} {\bibfnamefont {P.}~\bibnamefont
  {Giannozzi}}, \bibinfo {author} {\bibfnamefont {S.}~\bibnamefont {Baroni}},
  \bibinfo {author} {\bibfnamefont {N.}~\bibnamefont {Bonini}}, \bibinfo
  {author} {\bibfnamefont {M.}~\bibnamefont {Calandra}}, \bibinfo {author}
  {\bibfnamefont {R.}~\bibnamefont {Car}}, \bibinfo {author} {\bibfnamefont
  {C.}~\bibnamefont {Cavazzoni}}, \bibinfo {author} {\bibfnamefont
  {D.}~\bibnamefont {Ceresoli}}, \bibinfo {author} {\bibfnamefont {G.~L.}\
  \bibnamefont {Chiarotti}}, \bibinfo {author} {\bibfnamefont {M.}~\bibnamefont
  {Cococcioni}}, \bibinfo {author} {\bibfnamefont {I.}~\bibnamefont {Dabo}},
  \bibinfo {author} {\bibfnamefont {A.}~\bibnamefont {{Dal Corso}}}, \bibinfo
  {author} {\bibfnamefont {S.}~\bibnamefont {de~Gironcoli}}, \bibinfo {author}
  {\bibfnamefont {S.}~\bibnamefont {Fabris}}, \bibinfo {author} {\bibfnamefont
  {G.}~\bibnamefont {Fratesi}}, \bibinfo {author} {\bibfnamefont
  {R.}~\bibnamefont {Gebauer}}, \bibinfo {author} {\bibfnamefont
  {U.}~\bibnamefont {Gerstmann}}, \bibinfo {author} {\bibfnamefont
  {C.}~\bibnamefont {Gougoussis}}, \bibinfo {author} {\bibfnamefont
  {A.}~\bibnamefont {Kokalj}}, \bibinfo {author} {\bibfnamefont
  {M.}~\bibnamefont {Lazzeri}}, \bibinfo {author} {\bibfnamefont
  {L.}~\bibnamefont {Martin-Samos}}, \bibinfo {author} {\bibfnamefont
  {N.}~\bibnamefont {Marzari}}, \bibinfo {author} {\bibfnamefont
  {F.}~\bibnamefont {Mauri}}, \bibinfo {author} {\bibfnamefont
  {R.}~\bibnamefont {Mazzarello}}, \bibinfo {author} {\bibfnamefont
  {S.}~\bibnamefont {Paolini}}, \bibinfo {author} {\bibfnamefont
  {A.}~\bibnamefont {Pasquarello}}, \bibinfo {author} {\bibfnamefont
  {L.}~\bibnamefont {Paulatto}}, \bibinfo {author} {\bibfnamefont
  {C.}~\bibnamefont {Sbraccia}}, \bibinfo {author} {\bibfnamefont
  {S.}~\bibnamefont {Scandolo}}, \bibinfo {author} {\bibfnamefont
  {G.}~\bibnamefont {Sclauzero}}, \bibinfo {author} {\bibfnamefont {A.~P.}\
  \bibnamefont {Seitsonen}}, \bibinfo {author} {\bibfnamefont {A.}~\bibnamefont
  {Smogunov}}, \bibinfo {author} {\bibfnamefont {P.}~\bibnamefont {Umari}}, \
  and\ \bibinfo {author} {\bibfnamefont {R.~M.}\ \bibnamefont {Wentzcovitch}},\
  }\href {http://www.quantum-espresso.org} {\bibfield  {journal} {\bibinfo
  {journal} {J. Phys.}\ }\textbf {\bibinfo {volume} {21}},\ \bibinfo {pages}
  {395502} (\bibinfo {year} {2009})}\BibitemShut {NoStop}%
\bibitem [{\citenamefont {Troullier}\ and\ \citenamefont
  {Martins}(1991)}]{PhysRevB.43.1993.Troullier.Martins.pz.n.nc.pseudot.1991}%
  \BibitemOpen
  \bibfield  {author} {\bibinfo {author} {\bibfnamefont {N.}~\bibnamefont
  {Troullier}}\ and\ \bibinfo {author} {\bibfnamefont {J.~L.}\ \bibnamefont
  {Martins}},\ }\href {\doibase 10.1103/PhysRevB.43.1993} {\bibfield  {journal}
  {\bibinfo  {journal} {Phys. Rev. B}\ }\textbf {\bibinfo {volume} {43}},\
  \bibinfo {pages} {1993} (\bibinfo {year} {1991})}\BibitemShut {NoStop}%
\bibitem [{\citenamefont {Dal~Corso}(2012)}]{pseudopotentials}%
  \BibitemOpen
  \bibfield  {author} {\bibinfo {author} {\bibfnamefont {A.}~\bibnamefont
  {Dal~Corso}},\ }\href@noop {} {}\bibinfo {howpublished}
  {\url{http://qe-forge.org/gf/project/pslibrary/}} (\bibinfo {year}
  {2012})\BibitemShut {NoStop}%
\bibitem [{\citenamefont {Perdew}\ and\ \citenamefont
  {Zunger}(1981)}]{PhysRevB.23.5048-Perdew-Zunger-LDA}%
  \BibitemOpen
  \bibfield  {author} {\bibinfo {author} {\bibfnamefont {J.~P.}\ \bibnamefont
  {Perdew}}\ and\ \bibinfo {author} {\bibfnamefont {A.}~\bibnamefont
  {Zunger}},\ }\href {\doibase 10.1103/PhysRevB.23.5048} {\bibfield  {journal}
  {\bibinfo  {journal} {Phys. Rev. B}\ }\textbf {\bibinfo {volume} {23}},\
  \bibinfo {pages} {5048} (\bibinfo {year} {1981})}\BibitemShut {NoStop}%
\bibitem [{\citenamefont {Bl\"ochl}\ \emph {et~al.}(1994)\citenamefont
  {Bl\"ochl}, \citenamefont {Jepsen},\ and\ \citenamefont
  {Andersen}}]{PhysRevB.49.16223-Blochl-tetrahedron}%
  \BibitemOpen
  \bibfield  {author} {\bibinfo {author} {\bibfnamefont {P.~E.}\ \bibnamefont
  {Bl\"ochl}}, \bibinfo {author} {\bibfnamefont {O.}~\bibnamefont {Jepsen}}, \
  and\ \bibinfo {author} {\bibfnamefont {O.~K.}\ \bibnamefont {Andersen}},\
  }\href {\doibase 10.1103/PhysRevB.49.16223} {\bibfield  {journal} {\bibinfo
  {journal} {Phys. Rev. B}\ }\textbf {\bibinfo {volume} {49}},\ \bibinfo
  {pages} {16223} (\bibinfo {year} {1994})}\BibitemShut {NoStop}%
\bibitem [{\citenamefont {Karttunen}\ \emph {et~al.}(2011)\citenamefont
  {Karttunen}, \citenamefont {H\"{a}rk\"{o}nen}, \citenamefont {Linnolahti},\
  and\ \citenamefont {Pakkanen}}]{doi:10.1021/jp205676p.oma.Antti.mech.prop.}%
  \BibitemOpen
  \bibfield  {author} {\bibinfo {author} {\bibfnamefont {A.~J.}\ \bibnamefont
  {Karttunen}}, \bibinfo {author} {\bibfnamefont {V.~J.}\ \bibnamefont
  {H\"{a}rk\"{o}nen}}, \bibinfo {author} {\bibfnamefont {M.}~\bibnamefont
  {Linnolahti}}, \ and\ \bibinfo {author} {\bibfnamefont {T.~A.}\ \bibnamefont
  {Pakkanen}},\ }\href {\doibase 10.1021/jp205676p} {\bibfield  {journal}
  {\bibinfo  {journal} {J. Phys. Chem. C}\ }\textbf {\bibinfo {volume} {115}},\
  \bibinfo {pages} {19925} (\bibinfo {year} {2011})}\BibitemShut {NoStop}%
\bibitem [{\citenamefont {Nielsen}\ and\ \citenamefont
  {Martin}(1985)}]{PhysRevB.32.3792.Nielsen.Stresses.in.semic.1985}%
  \BibitemOpen
  \bibfield  {author} {\bibinfo {author} {\bibfnamefont {O.~H.}\ \bibnamefont
  {Nielsen}}\ and\ \bibinfo {author} {\bibfnamefont {R.~M.}\ \bibnamefont
  {Martin}},\ }\href {\doibase 10.1103/PhysRevB.32.3792} {\bibfield  {journal}
  {\bibinfo  {journal} {Phys. Rev. B}\ }\textbf {\bibinfo {volume} {32}},\
  \bibinfo {pages} {3792} (\bibinfo {year} {1985})}\BibitemShut {NoStop}%
\bibitem [{\citenamefont {San-Miguel}\ \emph {et~al.}(2002)\citenamefont
  {San-Miguel}, \citenamefont {M\'elinon}, \citenamefont {Conn\'etable},
  \citenamefont {Blase}, \citenamefont {Tournus}, \citenamefont {Reny},
  \citenamefont {Yamanaka},\ and\ \citenamefont
  {Iti\'e}}]{San-Miguel.et.al.2002.clath-PhysRevB.65.054109}%
  \BibitemOpen
  \bibfield  {author} {\bibinfo {author} {\bibfnamefont {A.}~\bibnamefont
  {San-Miguel}}, \bibinfo {author} {\bibfnamefont {P.}~\bibnamefont
  {M\'elinon}}, \bibinfo {author} {\bibfnamefont {D.}~\bibnamefont
  {Conn\'etable}}, \bibinfo {author} {\bibfnamefont {X.}~\bibnamefont {Blase}},
  \bibinfo {author} {\bibfnamefont {F.}~\bibnamefont {Tournus}}, \bibinfo
  {author} {\bibfnamefont {E.}~\bibnamefont {Reny}}, \bibinfo {author}
  {\bibfnamefont {S.}~\bibnamefont {Yamanaka}}, \ and\ \bibinfo {author}
  {\bibfnamefont {J.~P.}\ \bibnamefont {Iti\'e}},\ }\href {\doibase
  10.1103/PhysRevB.65.054109} {\bibfield  {journal} {\bibinfo  {journal} {Phys.
  Rev. B}\ }\textbf {\bibinfo {volume} {65}},\ \bibinfo {pages} {054109}
  (\bibinfo {year} {2002})}\BibitemShut {NoStop}%
\bibitem [{\citenamefont
  {Hall}(1967)}]{PhysRev.161.756.Hall.Bulk.modulus.Si.experim.1967}%
  \BibitemOpen
  \bibfield  {author} {\bibinfo {author} {\bibfnamefont {J.~J.}\ \bibnamefont
  {Hall}},\ }\href {\doibase 10.1103/PhysRev.161.756} {\bibfield  {journal}
  {\bibinfo  {journal} {Phys. Rev.}\ }\textbf {\bibinfo {volume} {161}},\
  \bibinfo {pages} {756} (\bibinfo {year} {1967})}\BibitemShut {NoStop}%
\bibitem [{\citenamefont {Basile}\ \emph {et~al.}(1994)\citenamefont {Basile},
  \citenamefont {Bergamin}, \citenamefont {Cavagnero}, \citenamefont {Mana},
  \citenamefont {Vittone},\ and\ \citenamefont
  {Zosi}}]{PhysRevLett.72.3133-Basile-et.al.Si-lat.const.}%
  \BibitemOpen
  \bibfield  {author} {\bibinfo {author} {\bibfnamefont {G.}~\bibnamefont
  {Basile}}, \bibinfo {author} {\bibfnamefont {A.}~\bibnamefont {Bergamin}},
  \bibinfo {author} {\bibfnamefont {G.}~\bibnamefont {Cavagnero}}, \bibinfo
  {author} {\bibfnamefont {G.}~\bibnamefont {Mana}}, \bibinfo {author}
  {\bibfnamefont {E.}~\bibnamefont {Vittone}}, \ and\ \bibinfo {author}
  {\bibfnamefont {G.}~\bibnamefont {Zosi}},\ }\href {\doibase
  10.1103/PhysRevLett.72.3133} {\bibfield  {journal} {\bibinfo  {journal}
  {Phys. Rev. Lett.}\ }\textbf {\bibinfo {volume} {72}},\ \bibinfo {pages}
  {3133} (\bibinfo {year} {1994})}\BibitemShut {NoStop}%
\bibitem [{\citenamefont {San-Miguel}\ \emph {et~al.}(1999)\citenamefont
  {San-Miguel}, \citenamefont {K\'egh\'elian}, \citenamefont {Blase},
  \citenamefont {M\'elinon}, \citenamefont {Perez}, \citenamefont {Iti\'e},
  \citenamefont {Polian}, \citenamefont {Reny}, \citenamefont {Cros},\ and\
  \citenamefont {Pouchard}}]{San-Miguel.et.al.1999.clath-PhysRevLett.83.5290}%
  \BibitemOpen
  \bibfield  {author} {\bibinfo {author} {\bibfnamefont {A.}~\bibnamefont
  {San-Miguel}}, \bibinfo {author} {\bibfnamefont {P.}~\bibnamefont
  {K\'egh\'elian}}, \bibinfo {author} {\bibfnamefont {X.}~\bibnamefont
  {Blase}}, \bibinfo {author} {\bibfnamefont {P.}~\bibnamefont {M\'elinon}},
  \bibinfo {author} {\bibfnamefont {A.}~\bibnamefont {Perez}}, \bibinfo
  {author} {\bibfnamefont {J.~P.}\ \bibnamefont {Iti\'e}}, \bibinfo {author}
  {\bibfnamefont {A.}~\bibnamefont {Polian}}, \bibinfo {author} {\bibfnamefont
  {E.}~\bibnamefont {Reny}}, \bibinfo {author} {\bibfnamefont {C.}~\bibnamefont
  {Cros}}, \ and\ \bibinfo {author} {\bibfnamefont {M.}~\bibnamefont
  {Pouchard}},\ }\href {\doibase 10.1103/PhysRevLett.83.5290} {\bibfield
  {journal} {\bibinfo  {journal} {Phys. Rev. Lett.}\ }\textbf {\bibinfo
  {volume} {83}},\ \bibinfo {pages} {5290} (\bibinfo {year}
  {1999})}\BibitemShut {NoStop}%
\bibitem [{\citenamefont {Nilsson}\ and\ \citenamefont
  {Nelin}(1972)}]{PhysRevB.6.3777.Nilsson.dispersion.experim.Sialpha}%
  \BibitemOpen
  \bibfield  {author} {\bibinfo {author} {\bibfnamefont {G.}~\bibnamefont
  {Nilsson}}\ and\ \bibinfo {author} {\bibfnamefont {G.}~\bibnamefont
  {Nelin}},\ }\href {\doibase 10.1103/PhysRevB.6.3777} {\bibfield  {journal}
  {\bibinfo  {journal} {Phys. Rev. B}\ }\textbf {\bibinfo {volume} {6}},\
  \bibinfo {pages} {3777} (\bibinfo {year} {1972})}\BibitemShut {NoStop}%
\bibitem [{\citenamefont {M\'elinon}\ \emph {et~al.}(1999)\citenamefont
  {M\'elinon}, \citenamefont {K\'egh\'elian}, \citenamefont {Perez},
  \citenamefont {Champagnon}, \citenamefont {Guyot}, \citenamefont {Saviot},
  \citenamefont {Reny}, \citenamefont {Cros}, \citenamefont {Pouchard},\ and\
  \citenamefont {Dianoux}}]{Melinon-PDOS-clath-I-II.1999-PhysRevB.59.10099}%
  \BibitemOpen
  \bibfield  {author} {\bibinfo {author} {\bibfnamefont {P.}~\bibnamefont
  {M\'elinon}}, \bibinfo {author} {\bibfnamefont {P.}~\bibnamefont
  {K\'egh\'elian}}, \bibinfo {author} {\bibfnamefont {A.}~\bibnamefont
  {Perez}}, \bibinfo {author} {\bibfnamefont {B.}~\bibnamefont {Champagnon}},
  \bibinfo {author} {\bibfnamefont {Y.}~\bibnamefont {Guyot}}, \bibinfo
  {author} {\bibfnamefont {L.}~\bibnamefont {Saviot}}, \bibinfo {author}
  {\bibfnamefont {E.}~\bibnamefont {Reny}}, \bibinfo {author} {\bibfnamefont
  {C.}~\bibnamefont {Cros}}, \bibinfo {author} {\bibfnamefont {M.}~\bibnamefont
  {Pouchard}}, \ and\ \bibinfo {author} {\bibfnamefont {A.~J.}\ \bibnamefont
  {Dianoux}},\ }\href {\doibase 10.1103/PhysRevB.59.10099} {\bibfield
  {journal} {\bibinfo  {journal} {Phys. Rev. B}\ }\textbf {\bibinfo {volume}
  {59}},\ \bibinfo {pages} {10099} (\bibinfo {year} {1999})}\BibitemShut
  {NoStop}%
\bibitem [{\citenamefont {Biswas}\ \emph {et~al.}(2008)\citenamefont {Biswas},
  \citenamefont {Myles}, \citenamefont {Sanati},\ and\ \citenamefont
  {Nolas}}]{Biswas.Nolas:033535.Therm.prop.ofSi136.2008}%
  \BibitemOpen
  \bibfield  {author} {\bibinfo {author} {\bibfnamefont {K.}~\bibnamefont
  {Biswas}}, \bibinfo {author} {\bibfnamefont {C.~W.}\ \bibnamefont {Myles}},
  \bibinfo {author} {\bibfnamefont {M.}~\bibnamefont {Sanati}}, \ and\ \bibinfo
  {author} {\bibfnamefont {G.~S.}\ \bibnamefont {Nolas}},\ }\href {\doibase
  10.1063/1.2960580} {\bibfield  {journal} {\bibinfo  {journal} {J. Appl.
  Phys.}\ }\textbf {\bibinfo {volume} {104}},\ \bibinfo {eid} {033535}
  (\bibinfo {year} {2008})}\BibitemShut {NoStop}%
\bibitem [{\citenamefont {Fabian}\ and\ \citenamefont
  {Allen}(1997)}]{PhysRevLett.79.1885-Allen-gruneisen-amorph.-Si}%
  \BibitemOpen
  \bibfield  {author} {\bibinfo {author} {\bibfnamefont {J.}~\bibnamefont
  {Fabian}}\ and\ \bibinfo {author} {\bibfnamefont {P.~B.}\ \bibnamefont
  {Allen}},\ }\href {\doibase 10.1103/PhysRevLett.79.1885} {\bibfield
  {journal} {\bibinfo  {journal} {Phys. Rev. Lett.}\ }\textbf {\bibinfo
  {volume} {79}},\ \bibinfo {pages} {1885} (\bibinfo {year}
  {1997})}\BibitemShut {NoStop}%
\bibitem [{\citenamefont {Strauch}\ \emph {et~al.}(1996)\citenamefont
  {Strauch}, \citenamefont {Pavone}, \citenamefont {Nerb}, \citenamefont
  {Karch}, \citenamefont {Windl}, \citenamefont {Dalba},\ and\ \citenamefont
  {Fornasini}}]{Strauch-1996-436-ab-initio-MSD}%
  \BibitemOpen
  \bibfield  {author} {\bibinfo {author} {\bibfnamefont {D.}~\bibnamefont
  {Strauch}}, \bibinfo {author} {\bibfnamefont {P.}~\bibnamefont {Pavone}},
  \bibinfo {author} {\bibfnamefont {N.}~\bibnamefont {Nerb}}, \bibinfo {author}
  {\bibfnamefont {K.}~\bibnamefont {Karch}}, \bibinfo {author} {\bibfnamefont
  {W.}~\bibnamefont {Windl}}, \bibinfo {author} {\bibfnamefont
  {G.}~\bibnamefont {Dalba}}, \ and\ \bibinfo {author} {\bibfnamefont
  {P.}~\bibnamefont {Fornasini}},\ }\href {\doibase
  http://dx.doi.org/10.1016/0921-4526(95)00770-9} {\bibfield  {journal}
  {\bibinfo  {journal} {Physica B}\ }\textbf {\bibinfo {volume} {219–220}},\
  \bibinfo {pages} {436 } (\bibinfo {year} {1996})}\BibitemShut {NoStop}%
\bibitem [{\citenamefont {Dalba}\ \emph {et~al.}(1999)\citenamefont {Dalba},
  \citenamefont {Fornasini}, \citenamefont {Grisenti},\ and\ \citenamefont
  {Purans}}]{PhysRevLett.82.4240-Dalba-1999-Ge-EXAFS}%
  \BibitemOpen
  \bibfield  {author} {\bibinfo {author} {\bibfnamefont {G.}~\bibnamefont
  {Dalba}}, \bibinfo {author} {\bibfnamefont {P.}~\bibnamefont {Fornasini}},
  \bibinfo {author} {\bibfnamefont {R.}~\bibnamefont {Grisenti}}, \ and\
  \bibinfo {author} {\bibfnamefont {J.}~\bibnamefont {Purans}},\ }\href
  {\doibase 10.1103/PhysRevLett.82.4240} {\bibfield  {journal} {\bibinfo
  {journal} {Phys. Rev. Lett.}\ }\textbf {\bibinfo {volume} {82}},\ \bibinfo
  {pages} {4240} (\bibinfo {year} {1999})}\BibitemShut {NoStop}%
\end{thebibliography}%
\end{document}